\pdfoutput=1
\documentclass[iop,numberedappendix,twocolappendix]{emulateapj}
\usepackage{enumerate}
\usepackage{enumitem}
\usepackage{graphics}
\usepackage{amsmath}
\usepackage{graphicx}
\usepackage{hyperref}
\usepackage{color}
\usepackage{epstopdf}
\usepackage{natbib}
\usepackage{threeparttable} 
\usepackage{comment}

\def\bfc{}
\def\bluet{}

\newcommand{\fdeg}{.\!\!^\circ}

\newcommand{\kms}{\mbox{\,km$\,$s$^{-1}$}}

\newcommand{\msun}{\mbox{M$_{\sun}$}\,}
\newcommand{\mpcs}{\msun\,pc$^{-2}$}

\newcommand{\sfrdens}{\msun\,{\rm yr}$^{-1}$\,{\rm kpc}$^{-2}$}

\newcommand \pc {\,{\rm pc }}
\newcommand{\vlsr}{\mbox{$V_{\mbox{\tiny LSR}}$}}

\newcommand{\twCO}{\mbox{$^{12}$\rm CO}\,}
\newcommand{\thCO}{\mbox{$^{13}$CO}\,}

\newcommand{\hi}{H~{\scriptsize I}\ }
\newcommand{\hii}{H~{\scriptsize II}\ }

\newcommand{\gthree}{\mbox{MCC331--90}\,}
\newcommand{\PV}{$\ell - V$}
\newcommand{\past}{{\rm past}}

\shorttitle{Molecular cloud and star formation in the surrounding of the giant \hii region RCW~106}
\shortauthors{H. Nguyen, Q. Nguyen Luong, P. G. Martin, et al.}

\begin{document}

\title{ THE THREE-MM ULTIMATE MOPRA MILKY WAY SURVEY. II. CLOUD AND STAR FORMATION NEAR THE FILAMENTARY MINISTARBURST RCW~106}
\author{Hans Nguy$\tilde{\hat{\rm e}}$n\altaffilmark{1,2},
Quang Nguy$\tilde{\hat{\rm e}}$n Lu{\hskip-0.65mm\small'{}\hskip-0.5mm}o{\hskip-0.65mm\small'{}\hskip-0.5mm}ng\altaffilmark{1,3,4},
Peter G. Martin\altaffilmark{1},
Peter J. Barnes\altaffilmark{5, 6},
Erik Muller\altaffilmark{4}, 
Vicki Lowe\altaffilmark{7,8},
Nadia Lo\altaffilmark{9},
Maria Cunningham\altaffilmark{8},
Fr$\acute{\rm{e}}$d$\acute{\rm{e}}$rique Motte,\altaffilmark{10}
Balthasar Inderm\"{u}hle\altaffilmark{7},
Stefan N. O'Dougherty\altaffilmark{11},
Audra K. Hernandez\altaffilmark{12},
Gary A. Fuller\altaffilmark{13}
}

\altaffiltext{1}{Canadian Institute for Theoretical Astrophysics, University of Toronto, 60 St. George Street, Toronto, ON M5S~3H8, Canada}
\altaffiltext{2}{Max-Planck-Institut f\"{u}r Radioastronomie, Auf dem H\"{u}gel 69, 53121 Bonn, Germany}
\altaffiltext{3}{EACOA Fellow at NAOJ, Japan \& KASI, Korea}
\altaffiltext{4}{National Astronomical Observatory of Japan, Chile Observatory, 2-21-1 Osawa, Mitaka, Tokyo 181-8588, Japan}
\altaffiltext{5}{Astronomy Department, University of Florida, P.O. Box 112055, Gainesville, FL 32611, USA}
\altaffiltext{6}{School of Science and Technology, University of New England, NSW 2351, Australia}
\altaffiltext{7}{CSIRO Astronomy and Space Science, P.O. Box 76, Epping, NSW 1710, Australia}
\altaffiltext{8}{School of Physics, University of New West Wales, NSW 2052 Australia}
\altaffiltext{9}{Departamento de Astronom\'{i}a, Universidad de Chile, Camino El Observatorio 1515, Las Condes, Santiago, Casilla 36-D, Chile}
\altaffiltext{10}{Laboratoire AIM Paris-Saclay, CEA/IRFU - CNRS/INSU - Universit\'e Paris Diderot, Service d'Astrophysique, B\^at. 709, CEA-Saclay, F-91191, Gif-sur-Yvette Cedex, France}
\altaffiltext{11}{College of Optical Sciences, University of Arizona, 1630 E. University Blvd., P.O. Box 210094, Tucson, AZ 85721, USA}
\altaffiltext{12}{Astronomy Department, University of Wisconsin, 475 East Charter St., Madison, WI 53706, USA}
\altaffiltext{13}{Jodrell Bank Centre for Astrophysics, Alan Turing Building, School of Physics and Astronomy, University of Manchester, Manchester, M13 3PL., UK}

\altaffiltext{\dag}{\url{hnguyen@cita.utoronto.ca}} 

\date{Received 2014 July10; accepted 2015 February 9; published 2015 MM DD}
\begin{abstract}

We report here a study of gas, dust and star formation rates (SFRs) in the molecular cloud complexes (MCCs) surrounding the giant H$\,{\rm \scriptsize{II}}$ region RCW$\,$106 using $^{12}$CO and $^{13}$CO$\,$(1-0) data from the Three-mm Ultimate Mopra Milky way Survey (ThrUMMS) and archival data. We separate the emission in the Galactic Plane around $l=330^{\circ}$-$335^{\circ}$ and $b=-1^{\circ}$-$1^{\circ}$ into two main MCCs: the RCW$\,$106 (V$_{\rm LSR} = -48\,$km$\,$s$^{-1}$) complex and the MCC331-90(V$_{\rm LSR} = -90\,$km$\,$s$^{-1}$) complex. While RCW$\,$106 (M$\sim 5.9\times 10^{6}\,$M$_{\odot}$) is located in the Scutum-Centaurus arm at a distance of 3.6$\,$kpc, MCC331-90 (M$\sim 2.8\times 10^{6}\,$M$_{\odot}$) is in the Norma arm at a distance of 5$\,$kpc. Their molecular gas mass surface densities are $\sim220$ and $\sim130\,$M$_{\odot}$ pc$^{-2}$, respectively. For RCW$\,$106 complex, using the 21$\,$cm continuum fluxes and dense clump counting, we obtain an immediate past ($\sim$-0.2$\,$Myr) and an immediate future ($\sim$+0.2$\,$Myr) SFRs of $0.25_{-0.023}^{+0.09}\,$M$_{\odot},{\rm yr}^{-1}$ and $0.12\pm0.1 \,$M$_{\odot}\,{\rm yr}^{-1}$. This results in an immediate past SFR density of $9.5_{-0.9}^{+3.4}\,$M$_{\odot}\,{\rm yr}^{-1}\,{\rm kpc}^{-2}$ and an immediate future SFR density of $4.8_{-3.8}^{+3.8}\,$M$_{\odot}\,{\rm yr}^{-1}\,{\rm kpc}^{-2}$. As both SFRs in this cloud are higher than the ministarburst threshold, they must be undergoing a ministarburst event although burst peak has already passed. We conclude that this is one of the most active star forming complexes in the southern sky, ideal for further investigations of massive star formation and potentially shedding light on the physics of high-redshift starbursts.

\end{abstract}


\section{Introduction}
\label{sect:intro}

\begin{table*}[!tbhp]
\centering
\small
\caption{Observation parameters of the data used in this study}
\begin{tabular}{c|cccccc}
\hline
\hline
\noalign{\vskip 2pt}
Instrument &Tracer& Frequency& Ang. Res& Velocity &$\Delta v_{res}$ & $1\sigma$ rms\\
or Survey & &(GHz) & (\arcsec)& (km s$^{-1}$)& (km s$^{-1}$) \\
\noalign{\vskip 2pt}
\hline
\noalign{\vskip 2pt}
CfA&$^{12}$CO &115.27 & 510 &-165 to 165 &1.3 &0.1 K km s$^{-1}$\\
ThrUMMS &$^{12}$CO &115.27 &72  &-140 to 10 &0.33 &$\sim 3$ K km s$^{-1}$\\
ThrUMMS &$^{13}$CO & 110.20 &72  &-140 to 10 &0.34 &$\sim 1$ K km s$^{-1}$\\
SGPS & \hi & 1.420 &132  & - &- &1.6 K km s$^{-1}$ \\
SGPS & 21\,cm cont. &1.420 &132  & - &- &1 mJy beam$^{-1}$ \\
MIPS/\emph{Spitzer} & $24\,\mu$m& &6  & -& -& 0.01 MJy sr$^{-1}$\\
PACS/\emph{Herschel} & $160\,\mu$m& &12  & -&- & 0.08 MJy sr$^{-1}$\\
SPIRE/\emph{Herschel} & $500\,\mu$m& &37  & -& - &1.2 MJy sr$^{-1}$\\
\noalign{\vskip 2pt}
\hline
\end{tabular}
\label{data}
\end{table*}

Studying the earliest phases of star formation involves examining the
morphological and kinematic structure of molecular clouds and the
transformation between different states of materials. Massive star formation often occurs in
molecular cloud complexes (MCCs) with radius $\sim 70~$pc or more (e.g.,
W43; e.g.,
\citealt{nguyenluong11b,motte14}), Cygnus X with radius $\sim 80~$pc (e.g.,
\citealt{schneider06,motte07}), and 
{\bfc the Central Molecular Zone (CMZ) with radius  $\sim 180$~pc (e.g., \citealt{miyazaki00,jones12}).} 
MCCs associated with massive star formation reside mainly
along the mid-plane of the Galaxy, may be distant and heavily
obscured in the visible regime. Though not being obscured in the radio wavelength, Large-field radio surveys of MCCs generally yield data sets with a substantial field of view but
insufficient resolution to conduct case-by-case studies of particular
clouds.  This is particularly true for low resolution surveys such as
the carbon monoxide (CO) survey of the Milky Way
\citep{bronfman1989,dame2001}. While they catalog molecular clouds,
the structure and properties of each cloud cannot be fully
resolved. However, recent high-angular-resolution spectroscopic
surveys of the Galactic plane such as the Galactic Ring Survey (GRS;
\citealt{jackson06}) and the Three-mm Ultimate Mopra Milky way Survey
(ThrUMMS; \citealt{barnes13}) help to address this problem.

Of particular interest here is the MCC of gas surrounding the
bright giant \hii region RCW~106, which was discovered in the
H$\alpha$ emission line survey of the southern Milky Way
\citep{rodgers1960}. The giant \hii region RCW~106 hosts a cluster
with mass $\sim 10^3\,\msun$\ and a Lyman continuum photon emission of
$10^{50}$ s$^{-1}$, likely originating from dozens of O-type stars
($M>8~\msun$). It resides in the Scutum--Centaurus arm at a distance of $\sim 3.6$~kpc from the Sun \citep{russeil05}. This makes it an excellent
laboratory for testing molecular cloud formation mechanisms in
response to galactic dynamical processes such as spiral density waves.

Moreover,  {\bfc the parent RCW~106 MCC is also the site of ongoing star
  formation.}  
It is centered on $l \sim 333\degr$, $b \sim
-0\fdeg5$ and has a local standard of rest velocity $\vlsr \sim
-50$\,\kms\ \citep{bains06}.{\bfc Its substructures have been subject to a variety of
spectral line and continuum studies, progressively unravelling the
star formation and evolutionary history of the region.} Specifically, \citep{lynga64} and \citep{urqhart07}
measured the impact of UV emitted by various nearby OB stars 
that ionize the surrounding environment.
1.2~mm continuum maps revealed 95 clumps with masses ranging
from 40 to $10^{4}$~M$_\odot$, some of which have infrared (IR)
counterparts suggesting embedded star formation \citep{mooker04}.
Higher density tracer such as \thCO, CS, HCO$^{+}$, HCN, and HNC emission lines show prominent velocity features centered on the
\vlsr\ of RCW~106 that coincide with 1.2~mm dust clumps and are also
sites of massive star formation \citep{bains06,wong08,lo09}.
A recent NH$_{3}$ survey toward dense clumps of the RCW~106 complex
revealed a large sample of cold collapsing clumps that are potentially
forming stars \citep{lowe14}. 
 
{\bfc 
The two densest and most massive sites of ongoing star formation core are
MMS5 and MMS68 \citep{mooker04}. They are associated with the ultracompact
\hii (UCH{\scriptsize II}) regions G333.6--0.2
\citep{fujiyoshi05,fujiyoshi06} and IRAS~16164-5096, respectively.
High resolution adaptive optics observations with 0.\arcsec3\
resolution ($\sim 1000$~AU) combined with mid-IR spectroscopy indicate
the presence of mid to late-type O stars that might still be gaining
mass \citep{grave14}.}

\begin{figure*}[!tbhp]
\centering
\includegraphics[scale=0.65]{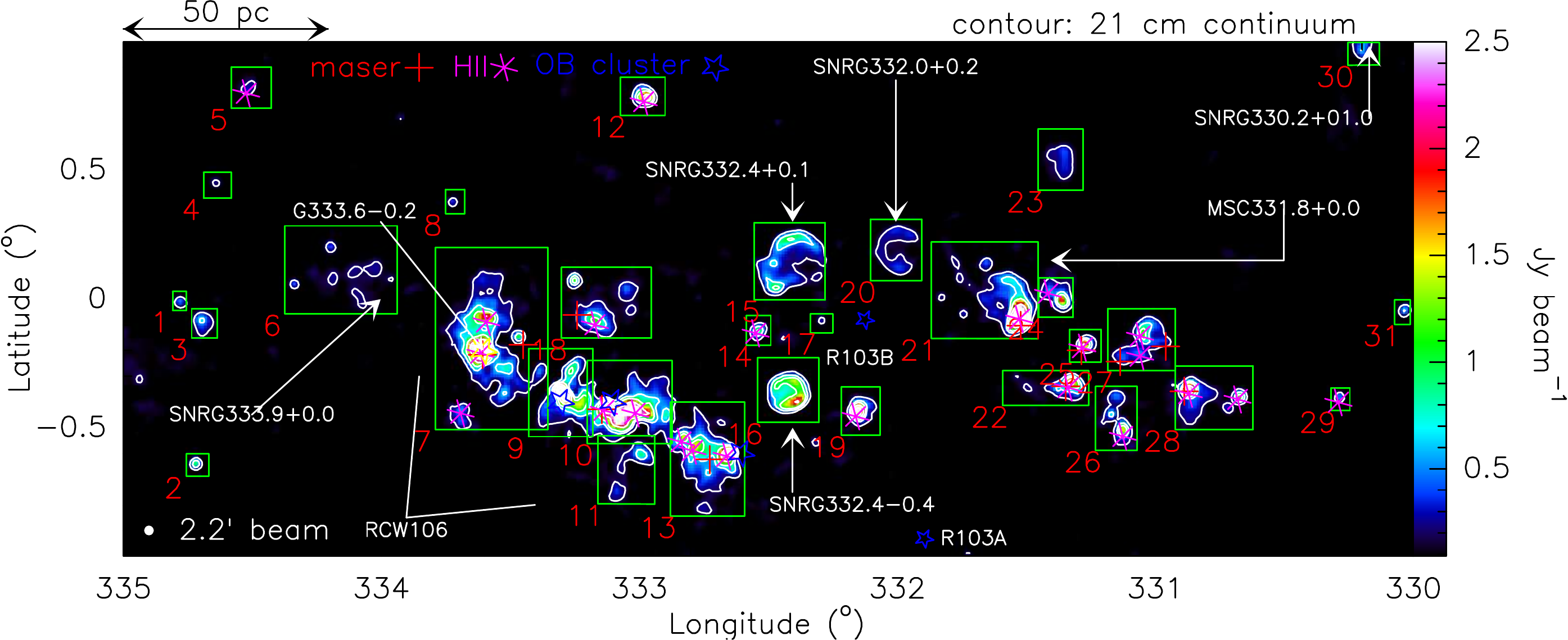}
\caption{21~cm continuum image (color scale) from the SGPS.  Contour
levels (white) are 0.2, 0.6, 1, and 2~Jy~beam$^{-1}$.  Within this region, we analysed continuum and molecular line data listed in Table~\ref{data} in 31 distinct boxes (green).
The linear ruler (top of diagram) assumes a distance of 3.6~kpc.
OH masers \citep{caswell1980} are plotted as red crosses, \hii regions
\citep{jones12a} as magenta asterisks, and prominent OB clusters as
blue stars. Further details are given in Appendix~\ref{AppendixA} and
Table~\ref{box_table}.}
\label{21cm_box}
\end{figure*}

{\bfc 
Previous studies focused mainly on the properties of gas
and dust at cloud scales of 0.1--10~pc (
\citealt{mooker04,bains06,wong08,lo09}), dense clump scales of 0.1--1~pc (\citealt{lo07,lo11}), or embedded O stars subparsec
scales of even smaller structure \citep{kumar13,grave14}.  In this paper we focus on the scale
of the cloud complex ($\sim50-100$~pc), which is the largest scale of molecular cloud in the Galaxy but the smallest scale that can be probed in nearby galaxies.  On this scale, RCW\,106 is a massive filamentary molecular cloud complex with an aspect ratio larger than 2: {see Figure \ref{fig:infrared}}. Our goal is to study the global
properties of the molecular clouds surrounding RCW~106 and establish a
relationship between the large scale structure of these clouds and their enhanced star formation. }

The various data sets used in this paper are described in
Section~\ref{sect:data}.
We present the two star-forming MCCs identified using mainly the \twCO emission in Section~\ref{sect:definition}.
The distances of these two cloud complexes are investigated in
Section~\ref{sect:distance_determination}.
Physical properties highlighting mass and mass surface density are
discussed in Section~\ref{sect:mass}.
The dynamical nature of the clouds are investigated in Section~\ref{sect:dyna}.
In Section~\ref{sect:sf}, we discuss the method of predcting the SFRs and the evidence that the RCW~106 complex can be quanlified as a ministarburst.

Finally, we summarize our findings in Section~\ref{nclusion}.

\section{Data}
\label{sect:data}

To investigate the properties of the region we used various continuum
and spectral line tracers, summarized in Table~\ref{data} and
discussed below.

\subsection{$^{12}$CO molecular line from the ThrUMMS survey}

The ThrUMMS survey \citep{barnes13} covers the Galactic plane between
$300\degr\le l \le 360\degr$ and $-1\degr \le b \le 1\degr$
in the
$^{12}$CO (1--0), $^{13}$CO (1--0), C$^{18}$O (1--0), and CN spectral
lines at frequencies of 115.27, 110.20, 109.78, and 113.60
GHz, respectively.
\footnote{
The data available to date for this study covered the range $-0\fdeg5 \le b \le 0\fdeg5$.}
The survey was conducted using the new Very Fast
Mapping (VFM) technique developed for use with the 22~m Mopra
telescope.\footnote{Operation of the Mopra radio telescope is made possible by funding from the National Astronomical Observatory of Japan, the University of New South Wales, the University of Adelaide, and the Commonwealth of Australia through CSIRO.}
Half-Nyquist sampling rates were obtained along the scanning direction
and between scan rows, yielding a 72\arcsec\ resolution. Compared to
the traditional on-the-fly mapping technique, the VFM technique is
capable of observing a larger area in the same amount of time if
Nyquist-sampled maps are not needed. As well as being more efficient,
it provides a nearly uniform sensitivity per unit area.  The data were
converted from antenna temperature $T_{\rm ant}$ to main-beam
temperature, $T_{\rm mb}$, by dividing $T_{\rm ant}$ by the
main-beam efficiencies $\eta_{\rm mb}$ of 0.42 for $^{12}$CO (1--0),
and of 0.43 for both $^{13}$CO (1--0) and C$^{18}$O (1--0).

\subsection{21~cm continuum and \hi line emission from the SGPS survey} 
\label{sgps_21cm_cont}

From the Southern Galactic Plane Survey (SGPS; \citealt{mcclure05}) we
extracted the \hi atomic line and 21~cm continuum data over the range
of $330\degr \leq l \leq 335\degr$ and $b= \pm 1\degr$. The survey
was conducted using the Australia Telescope Compact Array (ATCA)
interferometer and was complemented with the Parkes
64~m telescope (FWHM = 15\arcmin) for short spacings.  The observations were
performed simultaneously in a spectral line mode with 1024 channels
across a 4~MHz bandwidth centered at 1420~MHz and in a continuum mode
with 32 channels across a 128~MHz bandwidth centered at 1384~MHz
\citep{mcclure01,mcclure05}.  For our study, we used data which have a
FWHM 2\farcm2, line rms 1.6 K \kms, and continuum rms $<1$ mJy
beam$^{-1}$ (see Table~\ref{data}).

\subsection{Infrared data from Herschel and Spitzer}
\label{irsubmm}

We used 160 and 250\,$\mu$m images from the \emph{Herschel
Space Observatory} to examine more deeply the properties of the cold
dust near RCW~106.  The fields were observed as part of the
\emph{Herschel} Infrared Galactic Plane Survey (Hi-GAL;
\citealt{molinari10}) at $70/160\,\micron$ with the Photodetector
Array Camera and Spectrometer (PACS; \citealt{poglitsch10}) and at
$250/350/500\,\micron$ with the Spectral and Photometric Imaging
REceiver (SPIRE; \citealt{griffin10}).  The Hi-GAL data were taken in
parallel mode with a fast scanning speed of 60$\arcsec$ ${\rm
  s}^{-1}$.  The raw (level-0) data of each individual scan from both
PACS and SPIRE were calibrated and deglitched using HIPE\footnote{
HIPE is a joint development software by the \emph{Herschel} Science
Ground Segment Consortium, consisting of ESA, the NASA \emph{Herschel}
Science Centre, and the HIFI, PACS, and SPIRE consortia.} 
version 11.0. The SPIRE and PACS level-1 data were then
fed to version 18 of the {\sc Scanamorphos} software
package\footnote{
http://www2.iap.fr/users/roussel/herschel} \citep{roussel13} 
to regrid and create the final maps.  For our region, three Hi-Gal
fields of $2\fdeg5 \times 2\fdeg5$ were combined. The resultant
160 and 500\,$\mu$m images have angular resolutions of
12\arcsec\ and 37\arcsec, and $1\sigma$ rms of 0.08 and 1.0~MJy
sr$^{-1}$.

To trace the warm dust emission associated with high-mass star
formation, we used the $24\,\mu$m image from the Multi-band Imaging
Photometer for \emph{Spitzer} Galactic Plane Survey ({\it MIPSGAL};
\citealt{carey09}).

\section{Spatial and kinematic structure}
\label{sect:definition}

\subsection{21 cm continuum: defining the major structures}
\label{sect:21cm}

In the 21~cm continuum image (Fig.~\ref{21cm_box}) the giant \hii
region RCW~106 \citep{rodgers1960} is the largest
($1\fdeg2\times1\degr$) and brightest ($\sim 287$ Jy) area. RCW 106
is an elongated structure in $332\fdeg6<l<333\fdeg8$ and
$-0\fdeg8<b<0\fdeg1$, and angled roughly 60$^{\circ}$ with respect to
Galactic aast within the $l,b$ plane (hereafter referred to as the
``aastern" part).  
In the region $330\fdeg5<l<331\fdeg7$ and $-0\fdeg2<b<0\fdeg3$ there
is another similarly angled loose structure containing bright 21~cm
continuum sources (hereafter the ``western" part).  
Between these two complexes of thermal sources there is a
gap around $l \sim ~332^{\circ}$.  Projected in this gap is an
OB association R103B (332$\fdeg08$, $-0\fdeg08$) \citep{melnik95} and
also another further to the south, R103A ($331\fdeg89$, $-0\fdeg93$). These
are at spectroscopic distances of 3.22 and 3.00~kpc, respectively, and
so in the foreground. The eastern part is located at a distance of $\sim 3.6~$kpc and the western part is located at a distance of $\sim 5~$kpc

We have further divided the map in Fig.~\ref{21cm_box} into 31
subregions based on a contour corresponding to the $3\sigma$ noise
level above the local background.  Most of the sources are \hii
regions whose properties are classified according to their
morphologies in Appendix~\ref{AppendixA}. Additionally, there are four confirmed and two candidate supernova
remnants (SNRs) \citep{green09}.  For full details of each subregion
refer to Appendix~\ref{AppendixA} and the summary provided in
Table~\ref{box_table}.

In Fig.~\ref{21cm_box} we also mark the positions of OH masers
and \hii regions found in large
scale surveys \citep{caswell1980,jones12a}. The positions of masers and \hii regions
detected from radio recombination lines coincide with the bright radio continuum
sources help to distinguish substructures and to quantify the
cloud properties (see Sect.~\ref{sect:distance_determination}).

\subsection{CO and \hi line emission}
\label{sect:spectra}

\begin{figure}\centering
\includegraphics[width=275pt]{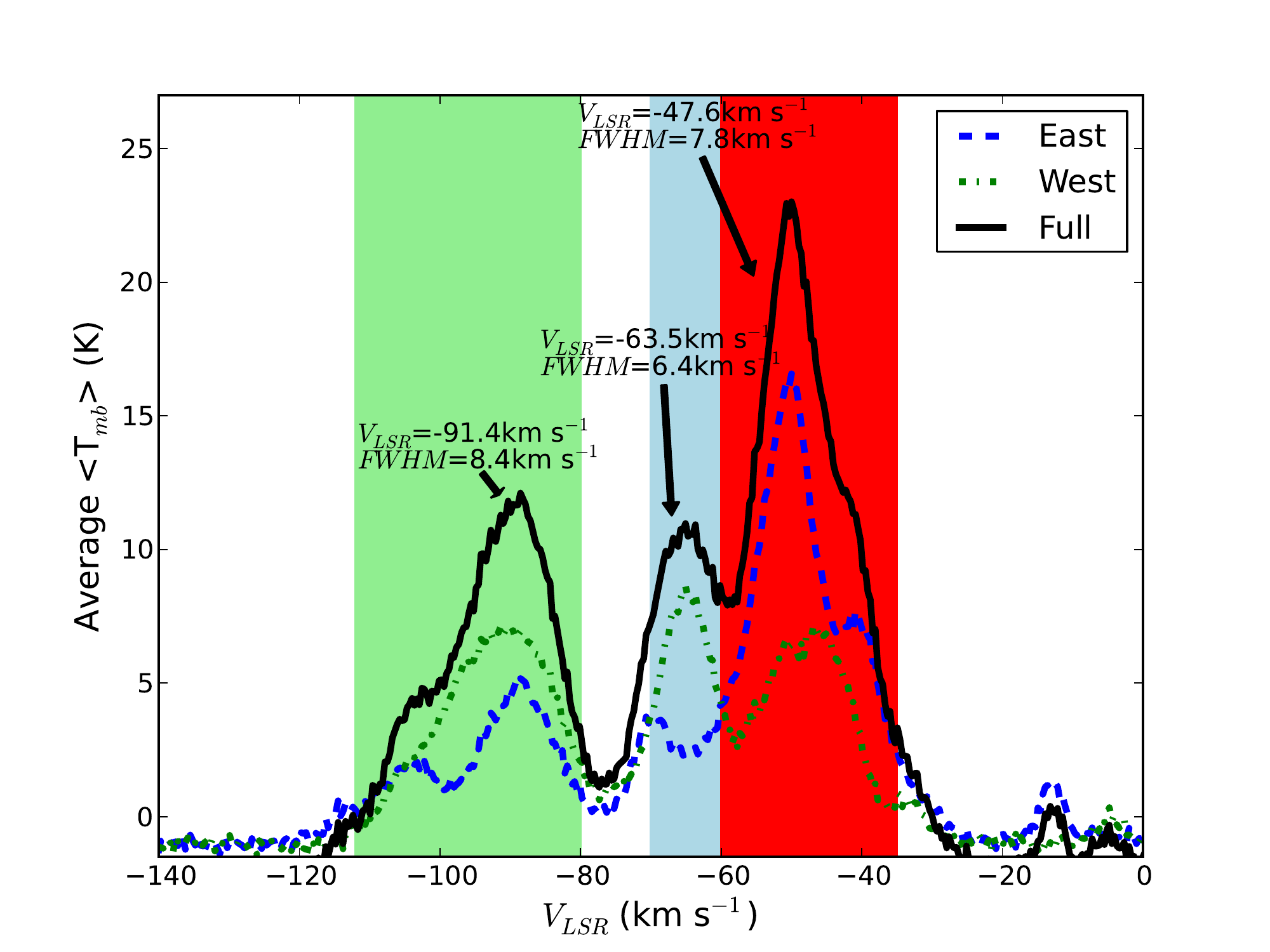}
\caption{The \twCO spectrum averaged over the entire region
($l=330\degr - 335\degr$, solid black), eastern part ($l=332\fdeg5 -
334\degr$, dotted green), and western part ($l=330\fdeg5 - 332\fdeg5$,
dashed blue) as defined in Fig.~\ref{21cm_box}.}
\label{int_12co_mains}
\end{figure}

To investigate the relative contributions of different clouds
superimposed along the line of sight, we first plot the average \twCO
spectrum of the entire region along with the distinctive spectra of
the two main structures, eastern and western, defined in Section
\ref{sect:21cm} (Fig. \ref{int_12co_mains}).  In the average spectrum
there are three distinct peaks within the velocity range $-140$ to
10\,\kms.  The spectra for both main structures exhibit significant
peaks at roughly the same velocities, in the ranges [$-112$, $-80$],
[$-72$, $-60$], and [$-60$, $-35$]\,\kms\ centered on $\vlsr = -91$,
$-63$, and $-48$\,\kms, respectively.  {\bfc These ranges are
  highlighted as green, blue, and red sectors in
  Fig.~\ref{int_12co_mains}.}  For the western region, the three
spectral maxima are roughly the same, on average $\sim 7$ K.  For the
eastern region, there are two peaks in the lowest velocity range, the
brighter one at $\vlsr \sim -55$\,\kms\ and the other at $\sim -45$
\,\kms\ (see Fig.~\ref{int_12co_mains}).  The peak at $\sim -55$ km
s$^{-1}$ coincides with the $^{13}$CO peak that \cite{bains06}
attributed to the cloud surrounding the RCW~106 cluster.

\begin{figure*}\centering
\includegraphics[scale=0.3]{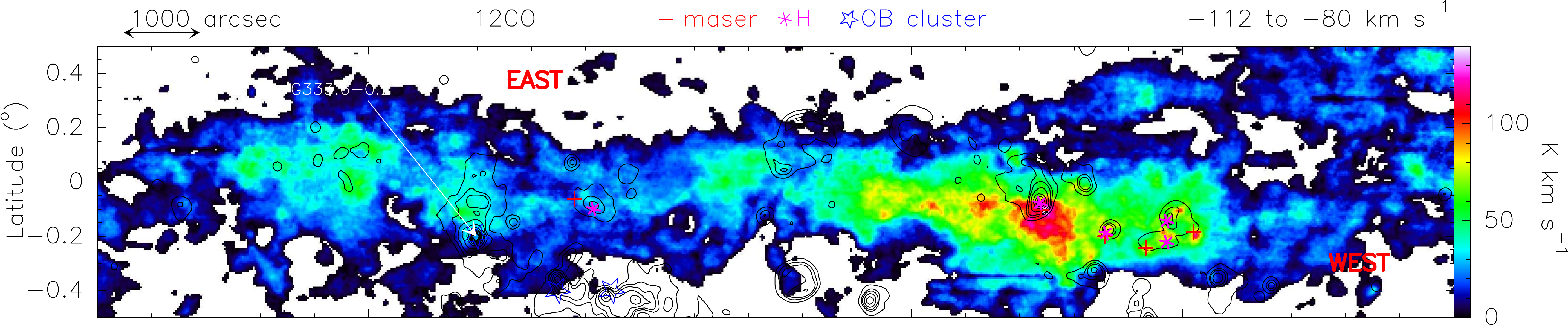}\\
\includegraphics[scale=0.3]{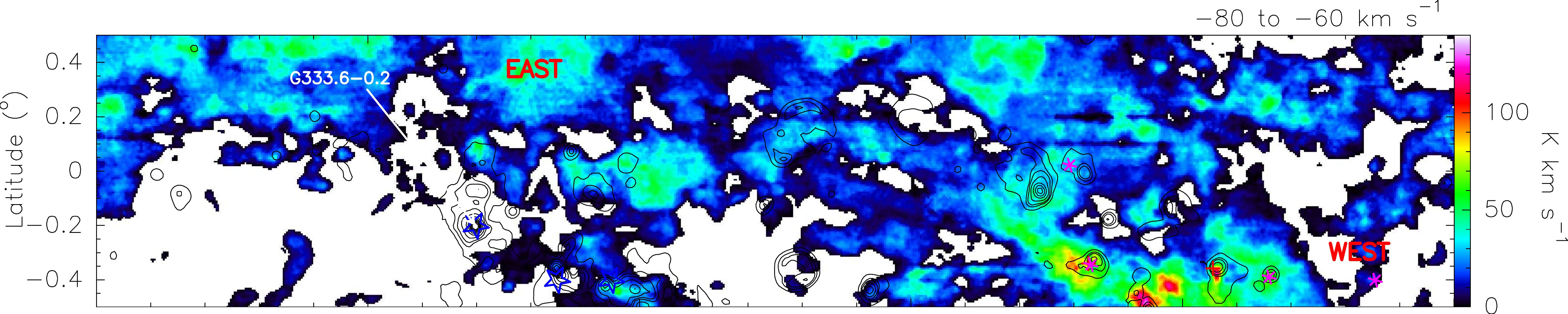}\\
\includegraphics[scale=0.3]{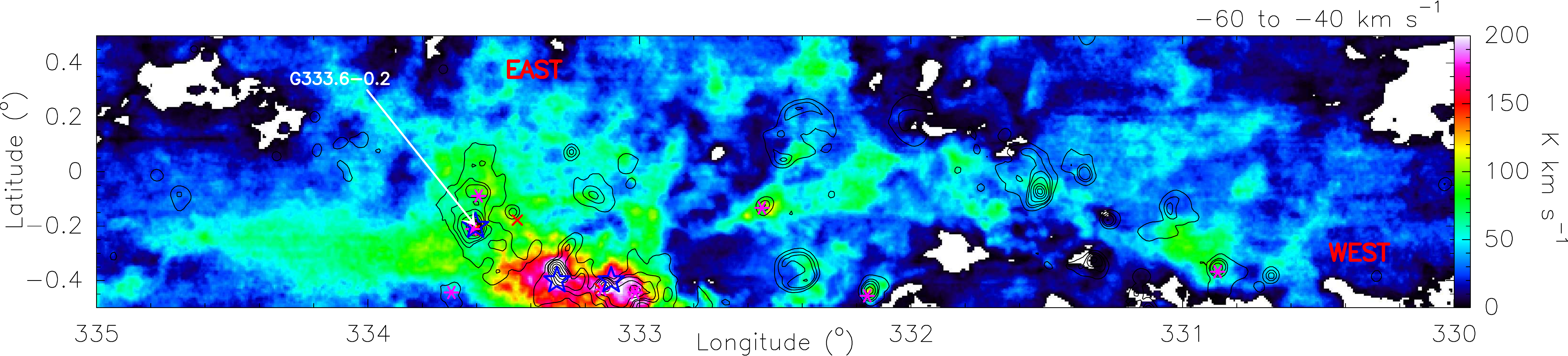}\\
\includegraphics[scale=0.3]{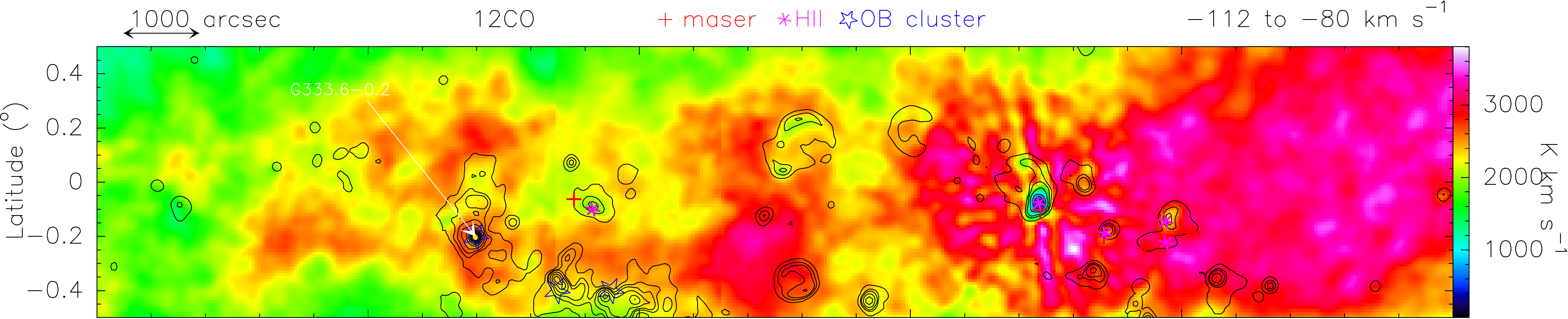}\\
\includegraphics[scale=0.3]{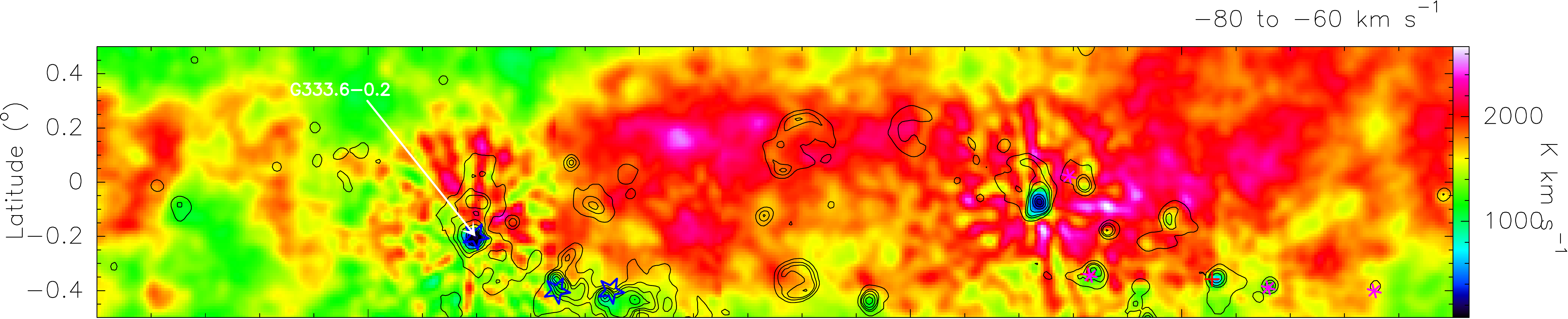}\\
\includegraphics[scale=0.3]{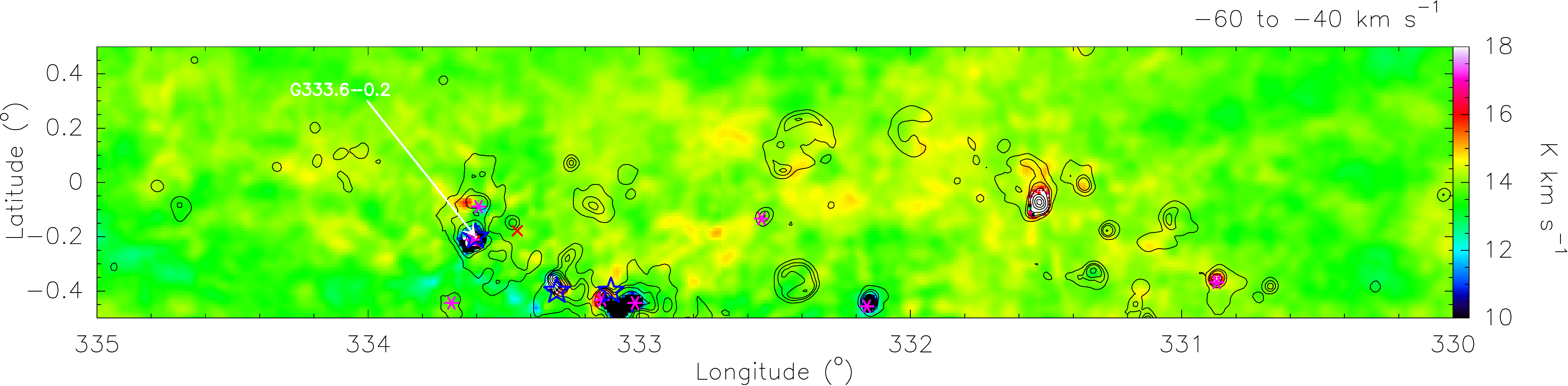}
\caption{{\bf Upper} -- \twCO integrated map for {\bfc (a)}: $-112$ to
$-80$\,\kms; {\bfc (b)}: $-80$ to $-60$\,\kms; {\bfc (c)}: $-60$ to
$-40$\,\kms. The black contours are from the 21 cm continuum emission
in Fig.~\ref{21cm_box} with levels of 0.2, 0.6, 1, and 2 Jy
beam$^{-1}$.  As in Fig.~\ref{21cm_box} positions of cataloged OH
masers and \hii regions are marked, but now only in the map with the
appropriate velocity range.  OB clusters are marked on all panels.
{\bf Lower} -- \hi integrated maps ({\bfc (d), (e), and (f)}) for the same
velocity ranges.  The black contours are the \twCO emission from the
corresponding panels above with levels in increments of 20 K\kms\ from
10 to 90 K\kms\ for {\bfc (d)} and {\bfc (e)} and 10 to 150 K\kms\ for
{\bfc (f)}.
}
\label{12co_-112_-80}
\end{figure*}

In Figs.~\ref{12co_-112_-80}a-c and d-f we display maps of the \twCO\
and \hi emission integrated over three velocity ranges:
[$-112$, $-80$], [$-80$, $-60$], and [$-60$, $-40$]\,\kms.
Additionally in Fig.~\ref{12co_-112_-80}, the positions of OH
masers and \hii regions mostly coincide both spatially and
spectrally with the molecular gas peaks.

Fig.~\ref{12co_-112_-80}a (corresponding to the green velocity range
in Fig.~\ref{int_12co_mains}) shows extended CO emission spanning the
entire longitude range. Its main peak is between 331\degr\ and
332\degr\ in longitude and extends just $0\fdeg2$ below the mid-plane;
it covers the western region and might extend to $l=333\fdeg4$.

The CO map integrated from $-80$ to $-60$\,\kms\ (the middle white and
blue velocity ranges in Fig.~\ref{int_12co_mains}), shown in
Fig.~\ref{12co_-112_-80}b, is dominated by the strongest emission from
the Western region, but is otherwise relatively free from emission
from the central plane (i.e., $b \sim 0\degr$).
On the other hand, the map shown in Fig.~\ref{12co_-112_-80}c, integrated from
$-60$ to $-40$\,\kms\ (the middle white and
red velocity ranges in Fig.~\ref{int_12co_mains}) is much brighter near the RCW~106 \hii region
and has a diffuse filamentary component in the western region.

The \hi emission in Figs.~\ref{12co_-112_-80}d-f does not appear to
correlate very well with either \twCO or 21 cm continuum emission,
indicating that star formation activity does not correlate with this
extended phase of the gas. However, in Figures~\ref{12co_-112_-80}d-e, \hi seems to form in the outer envelope of the MCCs.

\begin{figure*}\centering
\includegraphics[width=18cm,height=8cm]{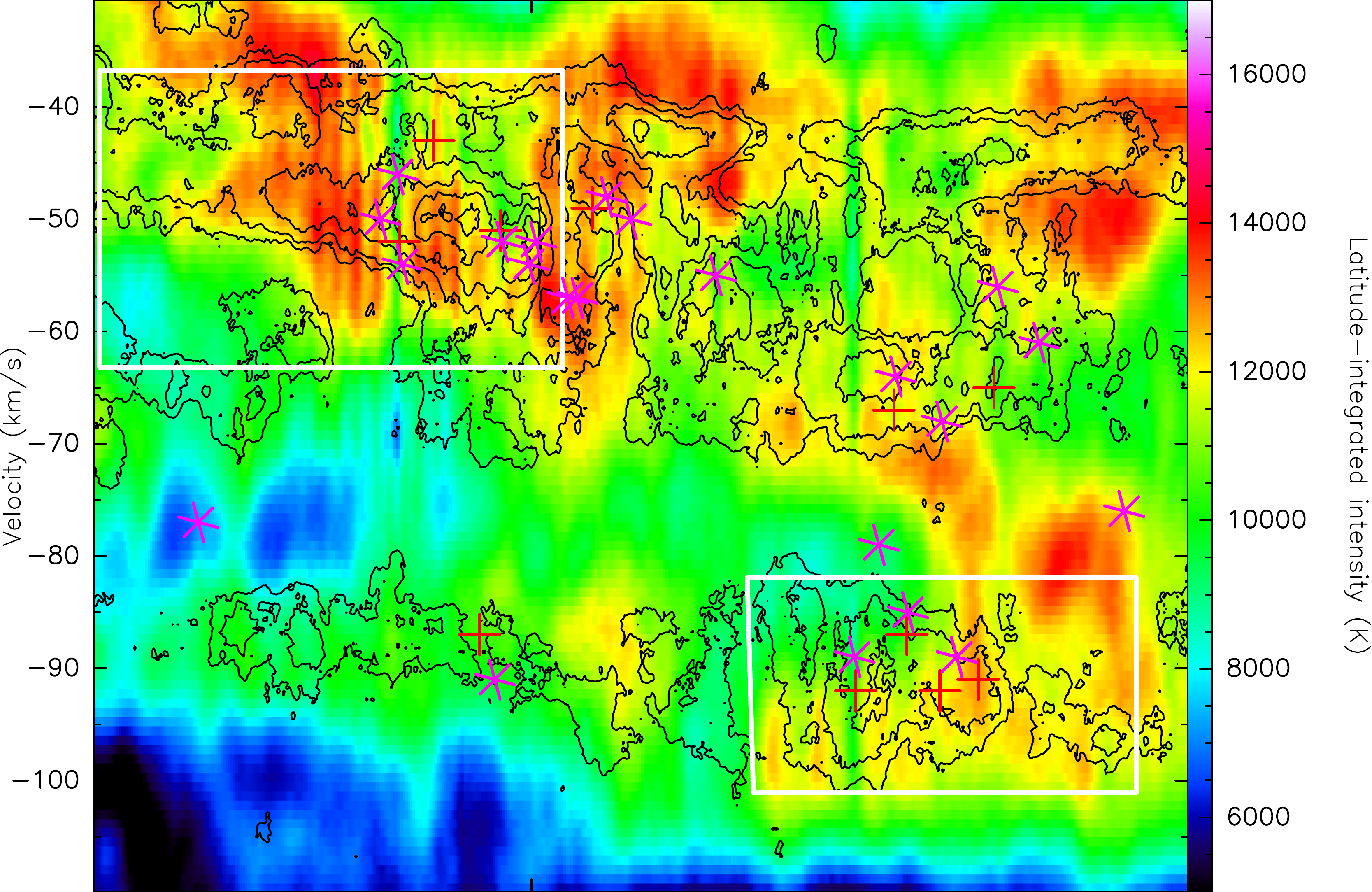}\\
\vskip 0.05cm
\includegraphics[width=18cm,height=8.5cm]{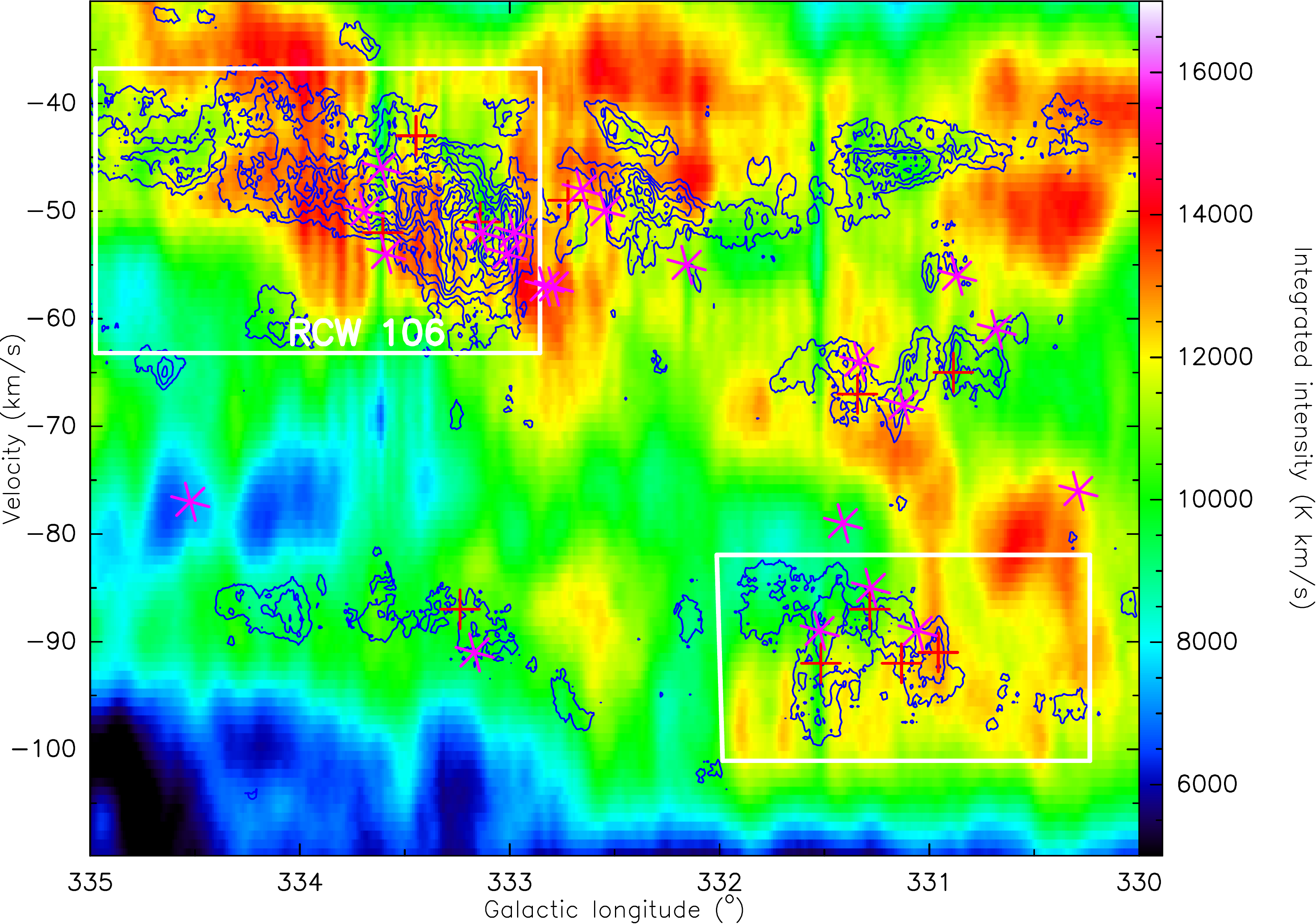}
\caption{\hi {\bf (color)}, \twCO {\bf (black contour, upper panel)}, and \thCO
{\bf (blue contour, lower panel)} position-velocity diagrams integrated from
$b=-0\fdeg5$ to $b=0\fdeg5$.
As in previous figures, red crosses represent OH masers and magenta
asterisks represent \hii regions. 
{\bfc 
The brightest 21~cm continuum sources in this region are at $l=331\fdeg5$ and $l=333\fdeg6$
(see Fig.~\ref{21cm_box} and Fig.~\ref{12co_-112_-80}).  
\hi absorption against these sources results in vertical bands of relatively
low (green) net \hi emission cutting through higher intensity (yellow, red) regions for 
the entire velocity range corresponding to foreground gas.
}
}
\label{fig:pv}
\end{figure*}

We have integrated the \twCO, \thCO, and \hi data cubes in Galactic
latitude (from $-0\fdeg5$ to $0\fdeg5$) to produce position-velocity
(\PV) diagrams (Fig.~\ref{fig:pv}). The main emission in the \PV\ maps
is from $-72$\,\kms\, to $-35$\,\kms\, as in the averaged spectra
(Fig.~\ref{int_12co_mains}).  Another feature of strong though less
prominent emission is from $-112$\,\kms\, to $-80$\,\kms\ at low
$\ell$ (western end).  This is detached from the lower velocity
structure supporting our division of the emission into two different
complexes: in the East is the RCW~106 MCC spanning the velocity
range from $-72$ to $-40$\kms, the West is 
{\bfc the complex 
{MCC 331.0+0.0 ($\vlsr
=-$90\,\kms)} (hereafter referred to with the informative name \gthree)}
spanning the velocity range from $-112$ to $-80$\,\kms.

\begin{figure*}[!tbhp]
\centering
$\begin{array}{c}
\includegraphics[scale=0.6]{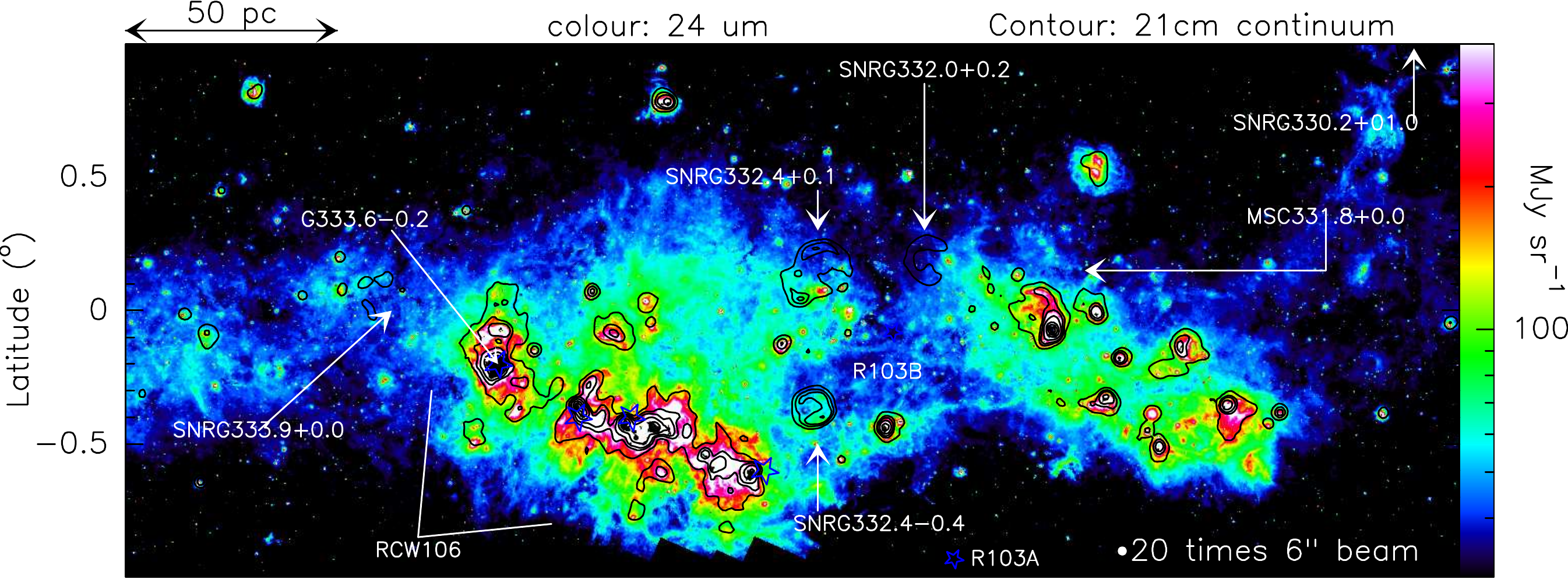}\\ 
\\
\includegraphics[scale=0.6]{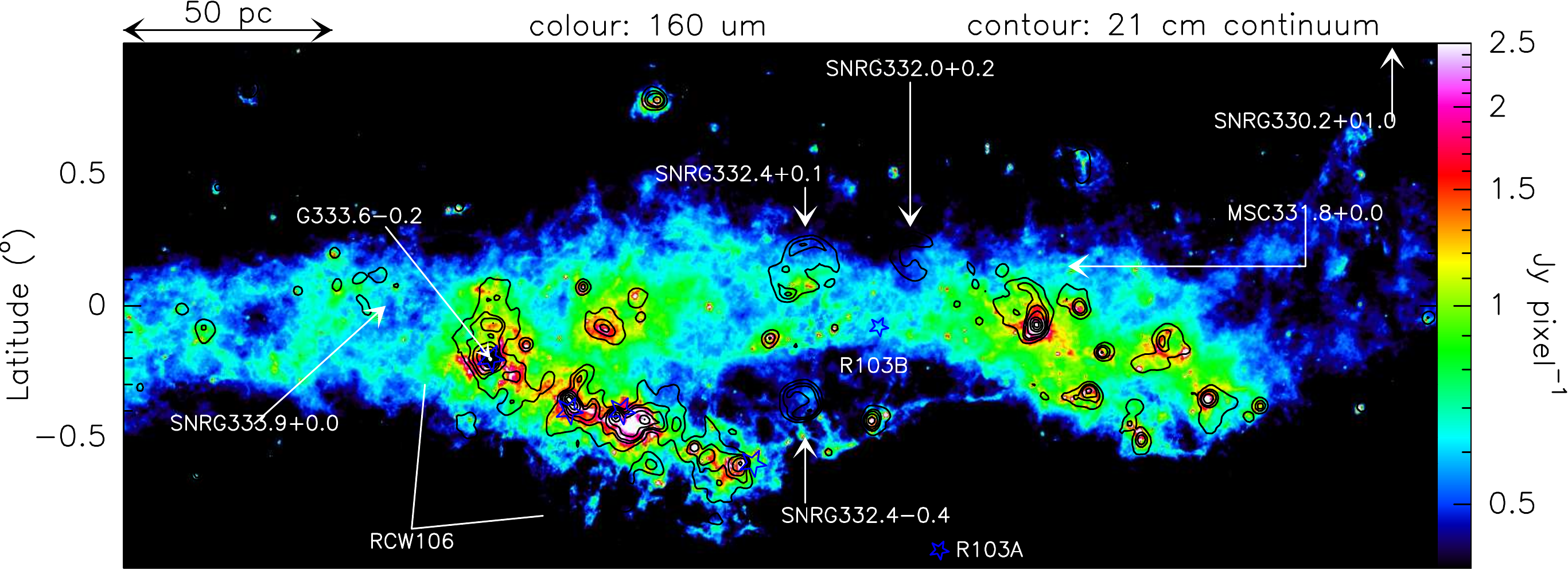}\\
\\
\includegraphics[scale=0.6]{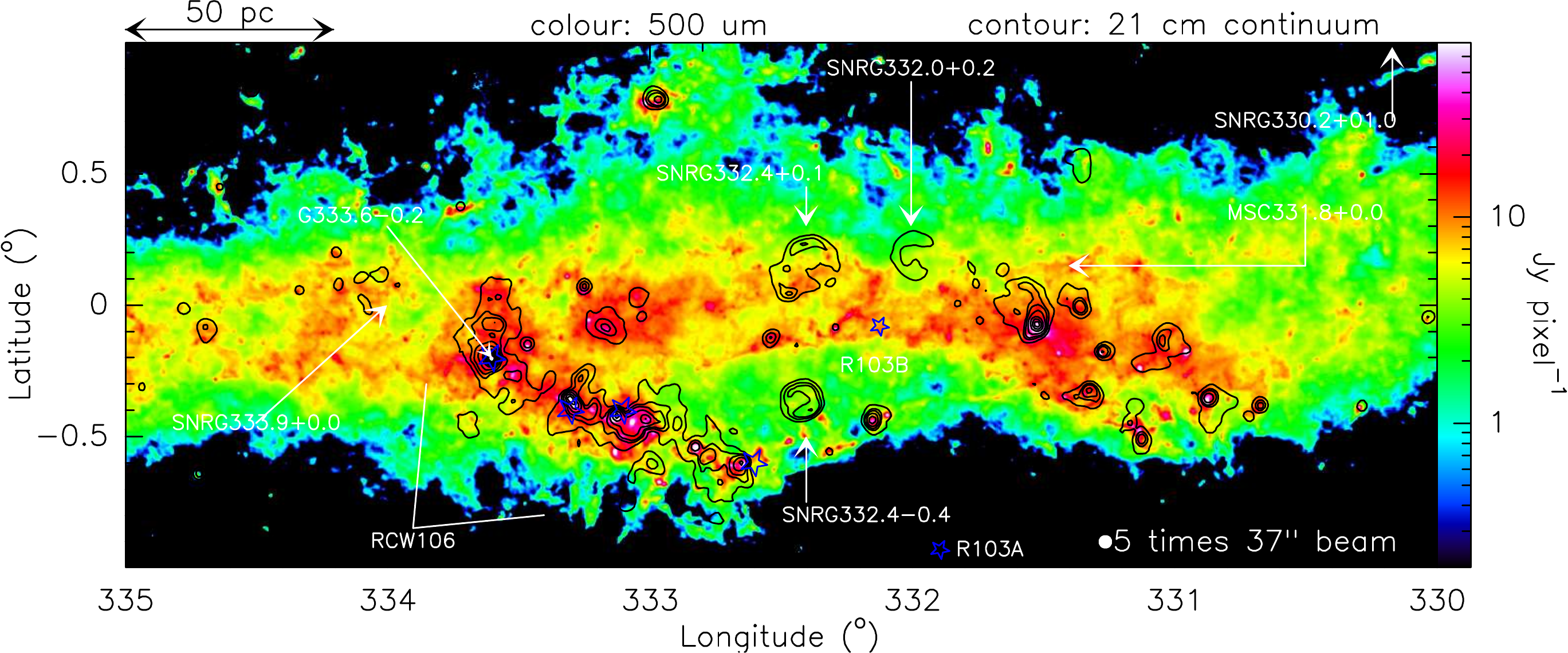}
\end{array}$
\caption{Dust emission {\bf (colour)} from \emph{Spitzer} 24\,$\mu$m ({\bfc
Top}), \emph{Herschel} 160\,$\mu$m ({\bfc Middle}), and
\emph{Herschel} 500\,$\mu$m ({\bfc Bottom}), overlaid with 21 cm
continuum emission contours (black). Blue stars mark the locations of
bright OB clusters.}
\label{fig:infrared}
\end{figure*}

\subsection{Insight from thermal dust emission}
\label{dust_emission}

Figure~\ref{fig:infrared} shows images of the thermal dust emission
obtained in the 
{\bfc mid-IR by \emph{Spitzer} and in the far-IR and submillimeter} 
by \emph{Herschel}
(Sect.~\ref{irsubmm}).\footnote{
Emission from Polycyclic Aromatic Hydrocarbons (PAHs) at $8\,\micron$
(not shown here) is also prominent in massive star forming regions
\citep{peeters04}.}  
These provide insight into the dust content, cloud morphology, and
radiation field.  In particular, data at $24\,\micron$
(Fig.~\ref{fig:infrared}a) allow us to trace local dust heating from
the UV photons from young OB stars.  Far-IR and submillimeter data
probe cooler dust as well, both in the cloud and outside, and away
from the influence of OB stars (Figures~\ref{fig:infrared}b-c).

The morphology of far-IR emission is generally similar to the CO
emission (Fig.~\ref{12co_-112_-80}) suggesting that the gas and dust
coexist within the MCCs.
{\bfc We also detect infrared dark clouds (IRDCs), seen as absorption in the
mid-IR data but as emission in the far-IR data. These IRDCs are
potential sites of massive star formation
\citep{hennebelle01,rathborne06, simon06b, peretto10c,
  nguyenluong11b}. 
} 
In the dust emission at all wavelengths we see two
main concentrations of dust at 331\degr\ and 333\degr\ whose detailed
morphology matches well with that in the angled structures seen in the
radio continuum and in \twCO. This correlation indicates that both the
21\,cm continuum and the dust heating are generated by the massive
stars in the RCW~106 and \gthree MCCs.
The \emph{Herschel} far-IR data also reveal finger-like protrusions at the
bottom of the structure linked to \gthree. 
{\bfc These features are plausibly formed
by radiation pressure from numerous OB stars in the local interstellar
medium (e.g., \citealt{krumholz09b,gritschneder10,tremblin13})}.

In summary, the dust emission data from 8--500\,\micron\, confirm that this region has two distinct MCCs
that are both extremely active in forming massive
stars, including hosting a young stellar cluster.

\section{Distances and location of the of RCW~106 and \gthree complexes}
\label{sect:distance_determination}

As shown in Fig.~\ref{int_12co_mains}, the RCW~106 molecular cloud
complex has two peaks centered at $-48$ and $-63$\,\kms, whereas the
\gthree molecular cloud complex has only one \vlsr\ peak at $\sim
-90\,\kms$.  We calculate the kinematic distances using the Galactic
rotation curve of \citet{reid2009}, following the approach of
\citet{romanduval09}. For RCW~106 and \gthree, we obtain near distances of
3.3--4.1~kpc and 4.6--6.0~kpc, respectively, in contrast to far
distances 10.2--12~kpc and 8.5--10~kpc, respectively. The far-distance
solutions are very unlikely. First, at a distance of 11\,kpc the MCC
dimension would be $\sim 200$~pc,　tripled the actual size calculated from the near solution. Second, the mass of RCW~106 would be a $\sim 10^{7}-10^{8}$\,\msun, as massive as the CMZ. Furthermore, at a Galactic latitude of -1\degr\, it would
imply that these MCCs are  $\sim 300$ pc off the Galactic mid-plane,
which
seems implausible even for the thick disc.
The near kinematic distance to the RCW~106 complex is also consistent
with the photometric distance estimates to the RCW~106 OB cluster of
3.6~kpc \citep{moises11}. 

The 11 OH
maser emission found by \cite{caswell1980} lie in the spatial and velocity range
of either the RCW~106 or \gthree complex (see Figs.~\ref{21cm_box},
\ref{12co_-112_-80} and \ref{fig:pv}), except for one source that lies
in the spatial range of RCW~106 with an offset of $\sim 40\,\kms$ in
\vlsr. The average near kinematic distances of the OH masers also coincide with those of
these two MCC, therefore confirm our argument for near kinematic distance. 
\citep{jones12a} resolved the distance ambiguities for the RCW~106 and \gthree
complexes using \hi absorption and found that more than 75\% have near distances. 
.  Among these, 15 are assigned the near distances and
four are uncertain. 

To be consistent with the measurements from
other methods in the literature (i.e., \citealt{bains06,moises11}), we
adopt 3.6~kpc as the distance to RCW~106.
We also conclude that separation into two main molecular structures is
physically meaningful, with both connected to signposts of massive
star formation such as OH masers or \hii regions. Hereafter, we adopt
a distance of 3.6~kpc to the RCW~106 complex and 5~kpc to the \gthree
complex.

\begin{figure}[!tbhp]
\centering
\includegraphics[scale=0.094]{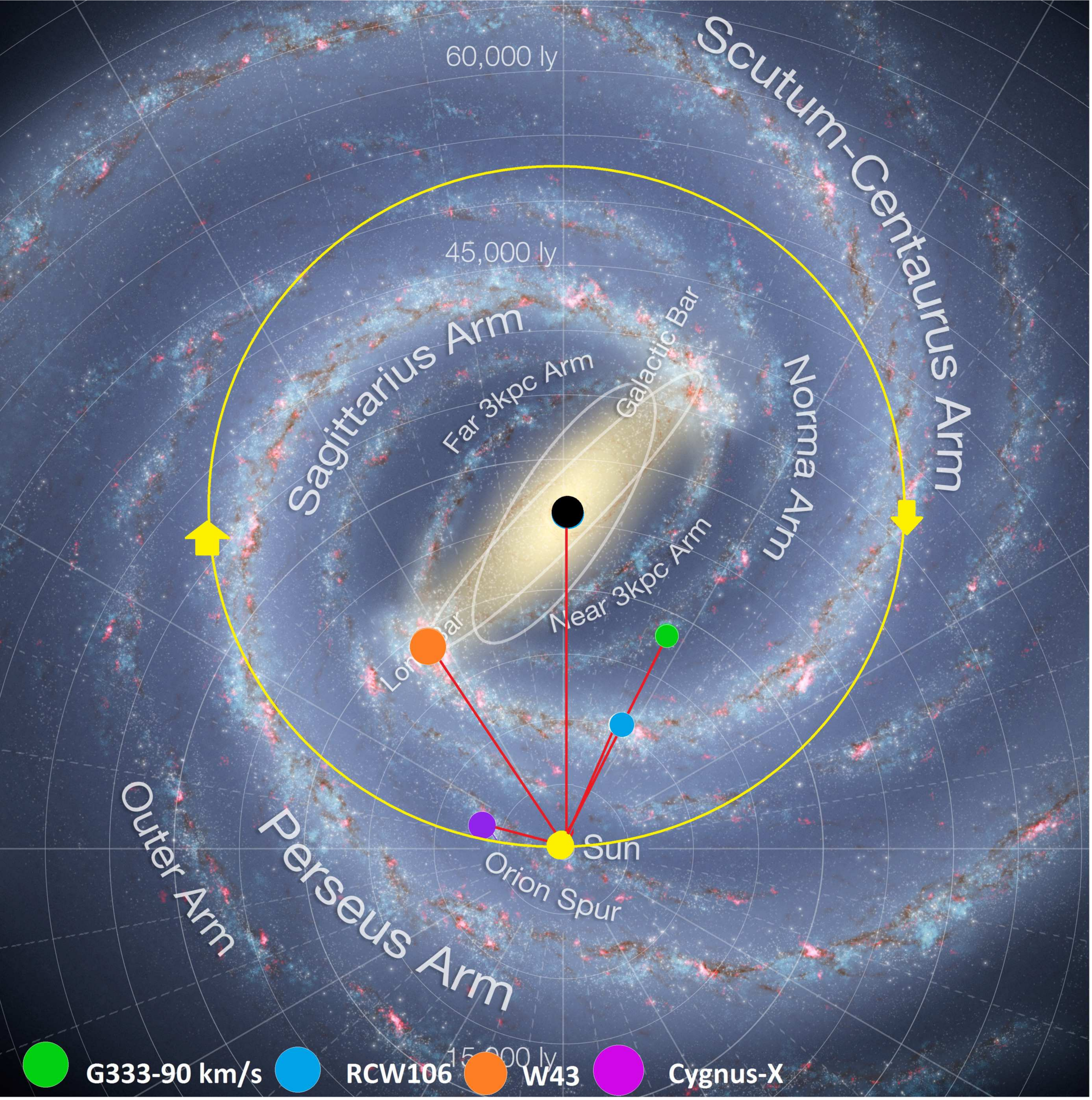}
\caption{Artist's rendition of the Milky Way seen face-on by Robert
Hurt of the \emph{Spitzer} Science Centre with advice from Robert
Benjamin from the University of Wisconsin--Whitewater. {\bfc Colored
dots mark the positions of GMCs, except for the black dot which marks
the Galactic center.} The RCW~106 complex is at $l=333\fdeg0$ and has
a distance of 3.6~kpc from the Sun, while \gthree is at $l=331\fdeg8$
and a distance of 5~kpc. Here it becomes obvious that the two GMCs are
separate and actually part of different spiral arms.}
\label{fig:galaxy_pos}
\end{figure}

With our calculated distances for these MCCs, we
can infer their locations in the Milky Way
(Fig.~\ref{fig:galaxy_pos}). Based on models of the Galaxy
\citep{georgelin1976,Rodriguez-Fernandez08,reid2009}, RCW~106 can be
placed in the nearby Scutum--Centaurus arm, the major arm of the
Galaxy, while \gthree is in the less prominent Norma arm.

The Scutum--Centaurus arm, a counterpart to the Perseus arm
(Fig.~\ref{fig:galaxy_pos}), is characterized by a high fraction of
dense gas \citep{sakamoto1997,russeil05} which may be the reason why
the most massive Young Massive Clusters (YMCs) such as Westerlund~1,
RSGC~1, RSGC~3 \citep{portegies-zwart10}, and the ministarburst W43
\citep{nguyenluong11} are found therein.  The \PV\
diagram of the CO and \hii emission shows a gradient toward the
RCW~106 \hii region (see upper dashed white line in Fig.~\ref{fig:pv})
that might reflect a velocity gradient along the Scutum--Centaurus arm
toward the RCW~106 complex.

Although the Norma arm, a counterpart to the Sagittarius arm
(Fig.~\ref{fig:galaxy_pos}), has less active star formation overall,
\gthree is known to host the luminous and efficient massive star
forming cloud G331.5--0.1 \citep{merello2013alma,merello2013}.

\section{Masses and mass surface densities of the RCW~106 and \gthree complexes}
\label{sect:mass}

\begin{table*}[!tbhp]
\scriptsize
\centering
\setlength{\tabcolsep}{3pt}
\caption{Properties of the two molecular cloud complexes in comparison
with other star-forming complexes}
\begin{threeparttable}
\begin{tabular}{l|cccccccccccc}
\hline
\hline
\noalign{\vskip 2pt}

Complex & $A_{\rm cloud}$\tnote{a} & $D$\tnote{b} &Vel. range\tnote{c} & $d$\tnote{d} & $M\tnote{e}$ & $\Sigma_{\rm gas}$\tnote{f}&$\sigma$\tnote{g}&$M_{\rm vir}$\tnote{h} & ${\alpha_{\rm vir}}$\tnote{i} & $\Sigma_{\rm SFR}$\tnote{j}\\

& (pc$^{2}$) & (pc) &(\kms) & (kpc) & (M$_\odot$) & (\mpcs) & (\kms) & (M$_\odot$) & & (\sfrdens) \\

\noalign{\vskip 2pt}
\hline
\noalign{\vskip 2pt}

RCW~106 &2.6 $\times 10^{4}$ &183&-80 to -40& 3.6 & 5.9$\pm1.77\times 10^{6}$&221.2&4.5&2.1$\times10^{6}$&0.35 & 0.15--0.3-{\bfc (0.3--2.3)}\tnote{k}\\

\gthree & 2.2 $\times 10^{4}$&167&-112 to -80& 5.0 &2.8$\pm0.84\times 10^{6}$&127.1&3.7&$1.4\times10^{6}$&0.5&- \\

\noalign{\vskip 2pt}
\hline
\noalign{\vskip 2pt}
\textless Gould Belt\textgreater\tnote{l} & 32.4\tnote{l} & 6.4\tnote{l}& - &0.27\tnote{l} & $3.0\times10^{3}$ &79.3\tnote{l}&-&-\\

W43\tnote{m} &1.5$\times 10^{4}$&$\sim 140$&80 to 100& 6.0 &7.1$\times 10^{6}$&473& 9.3 & 2.4$\times10^{7}$&0.3--0.5\tnote{m} & 0.65-{\bfc (6.5)}\tnote{k}\\

Cygnus X\tnote{m,n} &2.0$\times 10^{4}$&$\sim 160$&-10 to 20& 1.7 &5.0$\times 10^{6}$&250& 4.2 & 5.5$\times10^{6}$&0.9\\

CMZ\tnote{m,o}&9.6$\times 10^{4}$&$\sim 350$&-225 to 225& 8.5 &3.0$\times 10^{8}$&3125&-&-&-\\

\noalign{\vskip 2pt}
\hline
\end{tabular}
\vskip 2pt

\begin{tablenotes}
\scriptsize
\item[] $^a$Surface areas calculated at adopted distances.
$^b$Equivalent diameter $D=\sqrt{A_{\rm cloud}/\pi}$. 
$^c$Main velocity range of the structure.
$^d$Adopted distance.
$^e$Mass calculated from the CO emission using an $X$ of
$1.8\times10^{20}$ cm$^{-2}$ K$^{-1}$\kms\ except for Gould Belt clouds
being calculated from dust extinction map and for CMZ from thermal dust
emission.  
$^f$Mass surface density $\Sigma_{\rm gas}={M}/{A_{\rm cloud}}$.
$^g$Velocity dispersion $\sigma = {\Delta V_{\rm FWHM}}/{\sqrt{8\text{ln}2}}$.
$^h$Virial mass.
$^i$3D virial parameter described by $\alpha_{vir}={M_{vir}}/{M_{total}}$.
$^j$Star formation rate (SFR) density.
$^k$First numbers are the past SFR density and bold numbers in
brackets are the future SFR density (see Sect.~\ref{sect:sf}).
$^l$Average of 20 large molecular clouds from \emph{Spitzer} cores to
disks and Gould Belt surveys \citep{heiderman10}.
$^m$From \cite{nguyenluong11} with \emph{X} factor scaled to
$1.8\times10^{20}$ cm$^{-2}$ K$^{-1}$\kms.
$^n$From \cite{schneider06} using $^{13}$CO.
$^o$From \citet{dahmen98} using C$^{18}$O.
\end{tablenotes}
\end{threeparttable}
\label{tab:naming_structures_table}
\end{table*}

We obtain the mass from $W(\twCO)$, the \twCO spectrum integrated over
the velocity range $-80$ to $-40$\,\kms\ for RCW~106 and $-112$ to
$-80$~kms$^{-1}$ for \gthree.  The molecular hydrogen column density
is from $N_{{\rm H_2}} = X \times W(\twCO)$, with $X=1.8\times10^{20}$
cm$^{-2}$\,K$^{-1}$\kms\ \citep{dame2001}.  This $X$ factor is close
to the value of $1.9\times10^{20}$\,cm$^{-2}$\,K$^{-1}$\kms derived
for the Perseus molecular arm from the diffuse gamma-ray emission
\citep{fermi2010} and is lower than the value of $2.75\times10^{20}$
\,cm$^{-2}$\,K$^{-1}$\kms\ used for W43 and Cygnus X in
\cite{schneider06} and \cite{nguyenluong11}.  Although $X$ is
uncertain because of the optical depth, metallicity, and excitation
conditions, the mass estimates are probably accurate to a factor of
two. The mass $M_{\rm total}$ in an area $A_{\rm cloud}$ would then be $M_{\rm
  total} = N_{\rm H_2} \times A_{\rm cloud} \times \mu_{\rm H_2} m_{\rm H}$ where
$\mu_{\rm H_2}=2.8$ is the mean $H_{2}$ molecular weight and $m_{\rm
  H}$ is the H atomic mass.

We use the \twCO integrated intensity maps to define the extent of
each cloud and adopt a rectangular shape that covers the main extent
of the cloud (see Fig.~\ref{12co_-112_-80}) which is then used to calculate $A$ using the
assumed kinematic distance. The areas of the two clouds are
$2.6\times10^{4}$ pc$^{2}$ and $2.2\times10^{4}$ pc$^{2}$, which yield
effective diameters of 183 and 137 pc, respectively. However, we
integrate only within the contour $N_{\rm
  H_2}=5\times10^{22}$\,cm$^{-2}$ (corresponding to $W(\twCO) =
277$\,K\,\kms) to minimize the effects of the foreground and
background emission.
This yields masses $5.9\times10^{6}$\,\msun and
$2.8\times10^{6}$~\msun
and approximate surface densities 220 and 130\,\msun\,pc$^{-2}$ for
the RCW~106 and \gthree complexes, respectively. We estimate an
uncertainty of 30\% because of the background and foreground confusion
as well as optical depth. The properties are collected in
Table~\ref{tab:naming_structures_table}.

{\bfc 
The total clump mass deduced from measurements of $^{13}$CO
\citep{bains06}, mm continuum \citep{mooker04}, and far-IR
\citep{karnick2001} toward RCW~106 is only 10--30\% of our estimate,
for two reasons.  First, the previous measurements estimate only the
clump mass whereas we calculate the total gas mass which includes in
addition the more diffuse large-scale structure.}\footnote{
\bfc The ratio of total clump mass to the total gas mass can also be
considered as the clump formation efficiency, as elaborated further in
Section~\ref{sect:sf}.}
{\bfc 
Second, while these three studies focus in on the $\sim1$ square
degree area surrounding the RCW~106 cluster, we are studying a region
almost five times larger.}

\cite{heiderman10} investigated 20 low-mass star-forming clouds in the
Gould Belt region and found an average mass$\sim3000$\,$M_\odot$, area
32\,pc$^{2}$, and surface density of 80\,\msun pc$^{-2}$
(Table~\ref{tab:naming_structures_table}). Compared to these, the two
complexes studied here are $\sim 100$ times larger in area and yet
because the masses are so much higher the mass surface densities are
also higher. In reference to a typical massive star forming region such as
Cygnus~X, the mass surface density of RCW~106 is comparable, while
\gthree is about half.  But compared to the mass surface densities of the
extreme massive star forming regions such as W43 or CMZ molecular
cloud, the values here are lower still
(Table~\ref{tab:naming_structures_table}).

\section{Dynamics of RCW~106 and \gthree complexes}
\label{sect:dyna}

We fit a single Gaussian profile to the spectra of the
RCW~106 and \gthree MCCs and obtains FWHM widths of 10.6 and 8.7\,\kms\,
corresponding to a one-dimensional (1D) velocity dispersion $\sigma = \Delta V_{\rm FWHM}/\sqrt{\rm 8ln2}$ of 4.5 and 3.7\,\kms, respectively.
This value is within the typical range for
MCCs found in other observations (W43,
\citealt{nguyenluong11b}; Cygnus X, \citealt{schneider06}) or in
simulations of molecular clouds in spiral arms \citep{dobbs13}.

{\bluet First, to compare with other complexes, we estimate the gravitational instability criterium for the RCW\,106 complex using the spherical approximation and assuming that its effective radius $R_{\rm sph}$ (half the diameter $D$ in
Table~\ref{tab:naming_structures_table}) as:

\begin{equation}
\alpha_{\rm vir} = 5\times \frac{R_{\rm sph}\sigma_{\rm 1D}^{2}}{GM} = \frac{M_{vir}}{M_{\rm total}}\, .
\end{equation}

Both complexes have low {\it spherical virial parameters} ( 0.35 for RCW~106 and 0.5 for \gthree; see Table~\ref{tab:naming_structures_table}). 

However, since the global structure of RCW\,106 complex is filamentary with an aspect ratio $r=\frac{l}{\rm w}=\frac{\rm 180\,pc}{\rm 50\,pc}=3.6$, we use the gravitational instability criterium of an infinite-length isothermal gas cylinder (\citealt{ostriker64,inutsuka92}; see also \citealt{dibai58,ozernoi64}). This model defines a critical line mass (mass per unit length, $M_{\rm line,crit}$ solely on the sound speed $c_s$, above which the filamentary cloud is unstable against collapsing. For a large molecular cloud complex, the internal motion depends also on turbulence in addition to its thermal motion, therefore we assume that the apparent ''sound speed $c_s$'' equals the observed line width  $\Delta V$ (similar as in \citealt{dobashi14}).
We calculate the {\it filamentary virial parameter $\alpha_{\rm vir}^{\rm fil}$}, defined as:
\begin{equation}
\alpha_{\rm vir}^{\rm fil} = \frac{M_{\rm line,crit}}{M_{\rm line, gas}} = \frac{2c_{s}^{2}/G}{M/l}  = \frac{465 \Delta V}{M/l} \frac{[\rm \msun pc^{-1}]}{[\rm \msun pc^{-1}]} \, .
\end{equation}
The total gas mass $M$ is integrated over the RCW~106 and \gthree  filamentary clouds with length $l$ and width $w$. 
We obtain low  $\alpha_{\rm vir}^{\rm fil}$ of 0.15 for RCW~106 and 0.26 for \gthree, which confirm that these two complexes are indeed gravitationally unstable.
 
While the gravitationally unstable states of these clouds are certain, a more detailed investigation is needed to justify their origin.  
They might be caused by strong dynamical affects related to their locations spread
along the respective spiral arms (see Section~\ref{sect:distance_determination}) and/or feedback
from the vigorous burst of star formation
within (see Section~\ref{sect:sf}). 

}

\section{A burst of star formation in the RCW~106 complex}
\label{sect:sf}

Ly$\alpha$ continuum photons from young stars create \hii regions
quantifiable by free--free emission.  We find 30 \hii regions hosting O-
or B-type stars in the RCW~106 complex (Table~\ref{box_table}). We
checked the CO and \hi \vlsr\ of gas surrounding these \hii regions
and confirmed that they are in the velocity range of the RCW~106 complex.
We estimated the radio continuum flux density $S_{\nu}$ of each \hii
region by aperture photometry on the the 21\,cm map within the contour
defined by the 3$\sigma$ noise level (see Appendix~\ref{AppendixA}).
An average background level of $\sim 0.09$ Jy beam$^{-1}$ was derived
by averaging the intensity within a surrounding empty region.
{\bfc Following \cite{mezger67},} the number of Ly$\alpha$ continuum
photons powering an ionization-bounded region can be computed using
\begin{equation}
\frac{N_{\rm Ly\alpha}}{8.9\times10^{46}\,\text{s}^{-1}} = \frac{S_{\nu}}{\text{Jy}}   \left[\frac{\phantom{N} \nu \phantom{N}}{\text{GHz}}\right]^{0.1} \left[\frac{T_{e}}{10^{4}\,\text{K}}\right]^{-0.45}   \left[\frac{d}{\text{kpc}}\right]^{2} \, .
\label{SFR1}
\end{equation}
\noindent where $T_{e}=8000$~K \citep{wilson12} is the electron
temperature, $\nu=1.42$\,GHz is the observing frequency, and $d$ is
the distance to the \hii region.  The distance is 3.6~kpc for the
RCW~106 complex (see Sect.~\ref{sect:distance_determination}).

If we assume that a single main-sequence star dominates the
ionization,\footnote{
Of course, it is possible that the ionization is produced by a
combination of several stars of somewhat {\bfc later spectral type;} however,
usually only the brightest star dominates the output of ionizing
photons.}
we can then use $N_{\rm Ly\alpha}$ to obtain the spectral type
\citep{thompson1984,martins08}.  The estimated types are O7V or
earlier.

{\bfc 
Given that an O7 star with mass $\sim25\,\msun$ emits $N_{\rm Ly\alpha}
= 5\times10^{48}$\,s$^{-1}$ \citep{martins05}, we can estimate an
upper limit on the number of stars that have spectral type O7 or earlier in each \hii region of the RCW~106 cloud complex
(Table~\ref{box_table}); in total, there are up to 54 O7 stars
currently formed.
Assuming that the stellar masses in the RCW~106 complex follow the
Salpeter Initial Mass Function (IMF) $dN/dm= A \times m^{-2.35}$
\citep{salpeter55} and have a minimum cut-off mass of 0.08~\msun and
maximum cut-off mass of 50~M$_\odot$, the total stellar mass of RCW~106
will then be
\begin{equation}
M_{\rm *} =A\int m^{-\alpha}dm =  A \times \frac{{m_{\rm max}^{-0.35}}-m_{\rm min}^{-0.35}}{-0.35}\, .
\end{equation}

The normalization factor $A$ can be calculated from the total number
of O7 stars calculated above by the approximation that this is the
total number of all stars with masses in the range of 25 to 50\,\msun so
that
\begin{equation}
N_{m_{\rm l}}^{m_{\rm u}} =A \times \frac{{m_{\rm u}^{-1.35}}-m_{\rm l}^{-1.35}}{-1.35}\, .
\end{equation}

To gauge the uncertainty, we varied the upper mass limit to 100\,\msun
and allowed for a dispersion of $\pm$5 stars in the estimate of the
total number of O7 stars.  Finally, we obtain a value of $\sim7761_{-719}^{+2753}$ for the factor \emph{A}
and a total stellar mass of
$\sim48_{-4}^{+17} \times 10^3\,\msun$.
Replacing the Salpeter IMF by the Kroupa IMF \citep{kroupa01} lowers
the total stellar mass but it is within the calculated uncertainty.
}

\begin{figure*}[!tbhp]
\centering
\vskip -0cm
\hspace{-1.cm}
\includegraphics[scale=0.55]{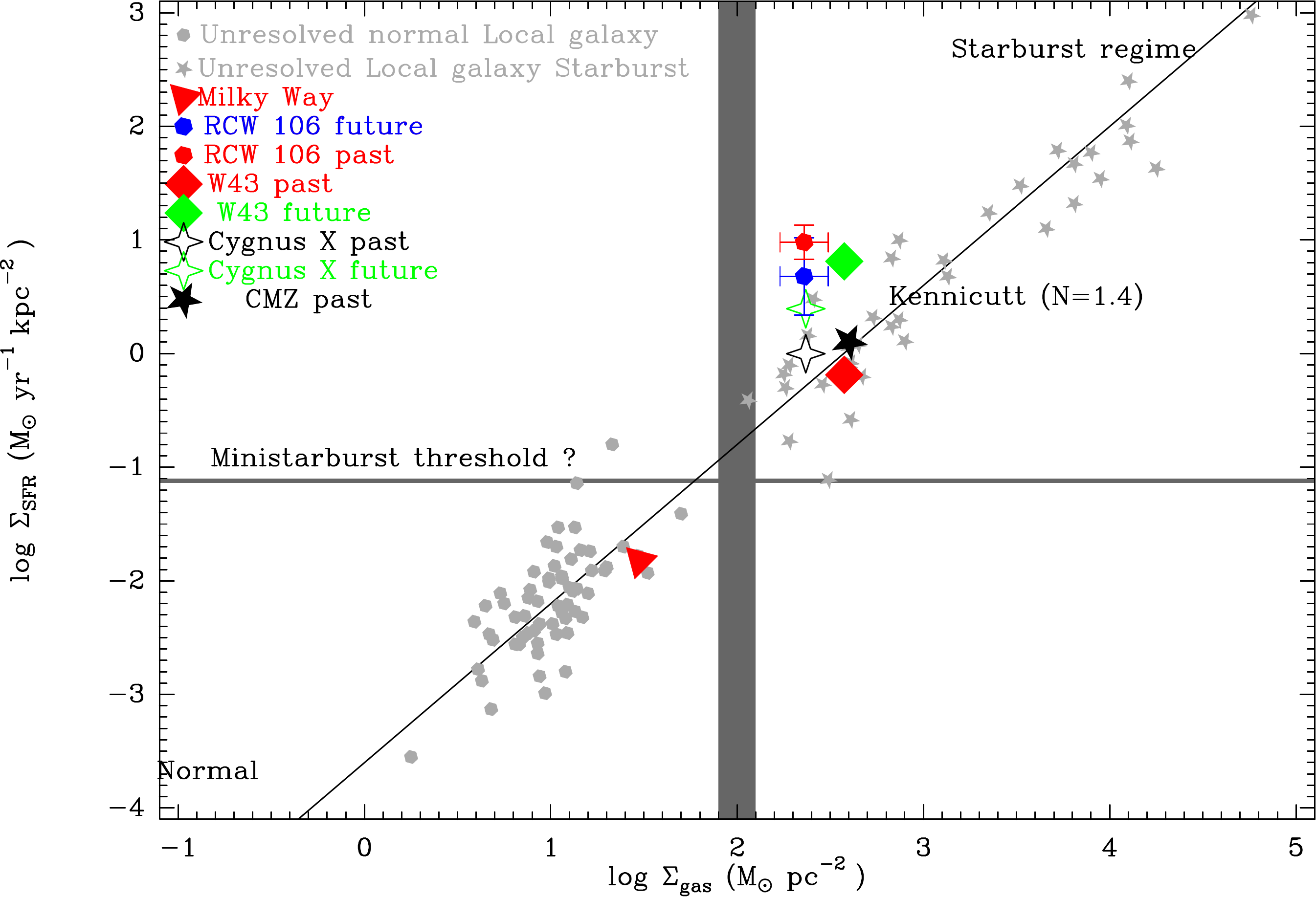}
\caption{The Schmidt-Kennicutt relation between the star formation
rate (SFR) density and gas surface density extending from normal
spiral galaxies to starbursts \citep{kennicutt98}.  Also plotted are
values for the average Milky Way, the immediate past and future of the
RCW~106 complex,
the past and future SFR density of the W43 complex, the  past and future SFR density of the Cygnus X complex, and  the past SFR density of the CMZ complex.}
\label{fig:sfrd_gd}
\end{figure*}

\subsection{The immediate past}
\label{impast}

{\bfc 
Because an O7V star with a mass of 25\,\msun has a lifetime of
$\sim2\times10^5$\,yr starting from its first appearance on the main-sequence diagram, we can calculate the star
formation efficiency $\epsilon_{\rm cloud}^{\past}$, the star
formation rate $SFR_{\rm cloud}^{\past}$, and the SFR density
$\Sigma_{\rm SFR, cloud}^{\past}$ of the entire RCW 106 cloud complex for this ``past"
time interval as
\begin{equation}
\epsilon_{\rm cloud}^{\past}  =  \frac{M_{\rm *}}{M_{\rm cloud}+M_{\rm *}} = 0.008_{-0.003}^{+0.004} \, ,
\end{equation}
\begin{equation}
SFR_{\rm cloud}^{\past}  = \frac{M_{\rm *}}{\rm 2\times 10^5~yr} =0.25_{-0.023}^{+0.09}\,\msun\,{\rm yr}^{-1} \, ,
\end{equation}
\begin{equation}
\Sigma_{\rm SFR, cloud}^{\past}  = \frac{SFR} {A_{\rm cloud}} =  9.5_{-0.9}^{+3.4}\,\msun\,{\rm yr}^{-1}\,{\rm kpc}^{-2} \, .
\end{equation}

with the total stellar mass $M_{\rm *}=4.8\times10^{4}\,\msun$, 
the total gas mass $ M_{\rm cloud} = 5.9\times10^{6}\,\msun$ 
calculated over an area of $A_{\rm cloud}=2.6\times10^{4}\,\text{\pc}^{2}$.

The estimated $\epsilon_{\rm cloud}^{\past}$ is in the middle of the
range for GMCs, 0.002--0.2, as derived from the ionizing flux of young
stars \citep{murray11}.  However, the SFR density of RCW~106 is high
and for its surface density of 220\,\msun\,pc$^{-2}$
(Sect.~\ref{sect:mass}) it is well above the trend in the Schmidt--Kennicutt
relation (see Fig.~\ref{fig:sfrd_gd}). This shows that RCW~106 has
been very active in forming massive stars during the last
$2\times10^5$ yr.
}

\subsection{The immediate future}
\label{imfuture}

The large amount of gas remaining in this complex suggests that the
burst of star formation activity in the RCW~106 molecular cloud complex
might not yet be finished. We investigated this by estimating the
future star-formation activity using the dense clump and core
populations.

First, {\bluet we make an assumption that the dense clump mass of $2.2-2.7\times 10^5\msun$ derived
from $^{13}$CO line emission \citep{bains06} or 1.2~mm continuum
emission \citep{wong08} represents the clump mass of the entire RCW 106 complex. This assumption is justified since most of the dense gas concentrates in the region considered by \cite{bains06} and \cite{wong08} (see for example Figure~\ref{fig:infrared}). }
We estimate the clump formation
efficiency, $\epsilon_{\rm cloud\rightarrow clump}$,\footnote{
This parameter is also known as the compactness of a molecular cloud
\citep{nguyenluong11}. }
as the ratio of the total clump mass to the total cloud mass
\citep{eden12,eden13,louvet14}:
\begin{equation}
\epsilon_{\rm cloud\rightarrow clump} = \frac{M_{\rm clump}}{M_{\rm cloud}} \sim ~0.042\pm0.013 \, . 
\end{equation}
The estimated $\epsilon_{\rm cloud\rightarrow clump}$ of the RCW~106
complex 
is about 10 times higher than that of the famous massive star-forming
region Cygnus~X and almost equal to that of W43 \citep{nguyenluong11}.
Thus the RCW~106 complex is forming massive clumps almost as
efficiently as the W43 star forming region despite the mass surface
density of RCW~106 being four times lower.

Secondly, a cloud's efficiency at converting mass into dense clumps
subsequently affects the star formation efficiency \citep{eden12,louvet14}. If
the dense clumps ($r\sim 1$\,pc) in the RCW~106 complex form
dense cores ($r\sim 0.1$~pc) with a mass transfer efficiency
$\epsilon_{\rm clump\rightarrow core}$, and dense cores will form
protostars with a mass transfer efficiency $\epsilon_{\rm
  core\rightarrow *}$, the future star formation efficiency of the
entire cloud will be
\begin{equation}
\epsilon_{\rm cloud\rightarrow *} = \epsilon_{\rm cloud\rightarrow clump} \times \epsilon_{\rm clump\rightarrow core} \times \epsilon_{\rm core\rightarrow *} \, . 
\end{equation}

If we adopt typical values $\epsilon_{\rm clump\rightarrow core}\sim
0.1-0.3$ \citep{parmentier13,louvet14} and $\epsilon_{\rm core\rightarrow
  *}\sim 0.3-0.8$ \citep{alves07,parmentier13,myers14}, we will have a
future $\epsilon_{\rm cloud\rightarrow *} \sim 0.004\pm0.004$.

Because the massive protostellar phase has a timescale of $\sim2\times
10^5$~yr \citep{motte07,russeil12}, similar or OB stars, we calculate
the ``future $2\times 10^5$~yr" (``future") 
SFR and SFR density as
\begin{equation}
\label{eq:starburstthreshold1}
SFR_{\rm cloud}^{\rm future} = \frac{\epsilon_{\rm cloud\rightarrow *} }{2\times 10^5 {\rm yr}} M_{\rm cloud} = 0.12\pm0.1 \,\msun\,{\rm yr}^{-1}
\end{equation}
\noindent and
\begin{equation}
\label{eq:starburstthreshold2}
\Sigma_{\rm SFR, cloud}^{\rm future} = \frac{SFR_{\rm cloud}^{\rm future}} {A_{\rm cloud}} = 4.8\pm3.8\, \msun\,{\rm yr}^{-1} {\rm kpc}^{-2} \, .
\end{equation}
This SFR density and the surface density also combine to place the
RCW~106 complex in the starburst regime of the Schmidt--Kennicutt
relation (\citealt{kennicutt98}; see Fig.~\ref{fig:sfrd_gd}).  It
should be possible to refine and confirm the future SFR density of the
RCW~106 complex by counting the massive dense cores identified using
the \emph{Herschel} data. This will be the topic of a subsequent paper.

\smallskip
These SFR densities are similar to the future SFR density of W43 and higher than that of Cygnus X, reinforcing the
conclusion that RCW~106 is undergoing a ministarburst at present. The
similarity between past and future SFR densities implies that this
ministarburst event is not yet finished but is in the declining phase. {\bluet It is also much higher than the SFR density of CMZ, the region of the galactic plane within a few
degrees of the Galactic centre, well-known for its inefficiency in star formation despite being 100 times more massive than RCW~106 complex \citep{immer12,kruijssen14}. }

While this is consistent with the large
mass reservoir still available in the RCW~106 complex, there are still
questions outstanding including what the final conversion efficiency
will be and its relationship to the virial parameter
(Section~\ref{sect:dyna}).

{\bluet 

\subsection{Uncertainties}
The calculations in Sections~\ref{impast} and \ref{imfuture} should involve the general uncertainties from the assumptions of the IMF, the timescales, and the mass transfer efficiencies. 
Several authors suggest that the IMF of the extreme star-formation environment is top-heavy \citep{stolte02,stolte05,chabrier14}. However, the validity of this statement is still largely debated  \citep{kim06,brandner08}. If we use an IMF with a steeper slope, the SFR likely increases by about 20\%.
We assume a burst period timescale of $2\times10^5$\,yr, which might have variations within a factor of two; but this is the most appropriate assumption since this is the measured timescale of an OB star \citep{martins05} and an MDC \citep{motte07}.   
The third uncertainty is the mass transfer efficiencies from clouds to clumps, to cores and to stars. Our assumptions of these quantities are the average results of the current state-of-the-art theoretical and observational studies, therefore the true values might vary at most 10\%--30\%.
We already take these three uncertainties into account in our calculations.

The uncertainties from measurements of Ly$\alpha$ photons and the clump mass are low since we focus only on the brightest regions in the radio continuum map (in case of estimating the Ly$\alpha$ photons) and in millimeter continuum or $^{13}$CO maps  (in case of estimating the clump mass). For the later case, the clump mass is verified with an independent measurement from NH$_{3}$ emission \citep{lowe14}.

}

\section{Conclusions}
\label{nclusion}

Using a combination of data sets, we have defined and characterized two
MCCs in the general direction of the RCW~106 OB
cluster, and we have labelled them the RCW~106 and \gthree complexes.
Their properties are summarized as follows

\begin{itemize}[leftmargin=0cm,itemindent=.4cm,labelwidth=\itemindent,labelsep=0cm,align=left]\itemsep0.5pt

\item The \vlsr\ for gas in the RCW~106 complex ranges between $-40$
  and $-80$\,\kms\, and the \vlsr\ for \gthree\, is between $-80$ and
  $-112$\,\kms. Both complexes have velocity dispersions comparable to
  that of Cygnus X ($\sigma$=4.2~\kms) but smaller in comparison to
  W43 ($\sigma$=9.3~\kms), which is right in the Galactic bar and a
  known turbulent area.

\item OH masers and \hii
  region studies confirmed the near kinematic distances of 3.6\,kpc for RCW~106 and 5\,kpc for \gthree.
  This places the RCW~106 molecular cloud
  complex in the Scutum--Centaurus arm and \gthree\,
  in the Norma arm.

\item The RCW~106 complex has a mass of $5.9\times10^6\, \msun$ and a
  surface density of $\sim 220\, \msun$pc$^{-2}$, whereas the \gthree
  complex has a mass of $2.8\times10^6\, \msun$ and surface density of
  $\sim 130\, \msun$pc$^{-2}$. These surface densities are higher
  compared to the average Gould Belt cloud ($\sim 70\,
  \msun$pc$^{-2}$) and the surface areas are $10^3$ times larger.

\item These two complexes are of comparable size ($d\sim$180~pc) but
  have less mass in comparison to other large GMCs such as W43 and
  Cygnus X.  Hence, the surface densities are smaller.

\item The virial parameters are greater than unity, indicating that
  they are gravitationally unbound.

\item Using the 21 cm continuum, we separated the region into 31
  subregions: 25 containing \hii regions and 6 containing SNRs.  We estimated
  that there are about 50 young O7V stars currently in the RCW~106 complex.

\item For the RCW~106 molecular cloud complex, we derive a past
  global star formation efficiency, SFR, and SFR density and estimate values to quantify star
  formation in the near future.  These values suggest that the RCW~106
  complex is undergoing a ministarburst event.

\end{itemize}

\vspace{-0.5cm}

\begin{acknowledgements}
\emph{Acknowledgements}: H.N. is grateful for a CITA summer student
internship. P.G.M. acknowledges support from the Canadian Space Agency
and the Natural Sciences and Engineering Research Council of
Canada. Nadia Lo's postdoctoral fellowship is supported by
CONICYT/FONDECYT postdoctorado under project No. 3130540.
\end{acknowledgements}

\bibliographystyle{aa} \bibliography{g333}

\begin{thebibliography}{94}
\expandafter\ifx\csname natexlab\endcsname\relax\def\natexlab#1{#1}\fi

\bibitem[{{Abdo} {et~al.}(2010){Abdo}, {Ackermann}, {Ajello}, {Baldini},
  {Ballet}, {Barbiellini}, {Bastieri}, {Baughman}, {Bechtol}, {Bellazzini},
  {Berenji}, {Bloom}, {Bonamente}, {Borgland}, {Bregeon}, {Brez}, {Brigida},
  {Bruel}, {Burnett}, {Buson}, {Caliandro}, {Cameron}, {Caraveo}, {Casandjian},
  {Cecchi}, {{\c C}elik}, {Chekhtman}, {Cheung}, {Chiang}, {Ciprini}, {Claus},
  {Cohen-Tanugi}, {Cominsky}, {Conrad}, {Dermer}, {de Palma}, {Digel}, {Silva},
  {Drell}, {Dubois}, {Dumora}, {Farnier}, {Favuzzi}, {Fegan}, {Focke},
  {Fortin}, {Frailis}, {Fukazawa}, {Funk}, {Fusco}, {Gargano}, {Gehrels},
  {Germani}, {Giavitto}, {Giebels}, {Giglietto}, {Giordano}, {Glanzman},
  {Godfrey}, {Grenier}, {Grondin}, {Grove}, {Guillemot}, {Guiriec}, {Harding},
  {Hayashida}, {Horan}, {Hughes}, {Jackson}, {J{\'o}hannesson}, {Johnson},
  {Johnson}, {Kamae}, {Katagiri}, {Kataoka}, {Kawai}, {Kerr}, {Kn{\"o}dlseder},
  {Kuss}, {Lande}, {Latronico}, {Lemoine-Goumard}, {Longo}, {Loparco}, {Lott},
  {Lovellette}, {Lubrano}, {Makeev}, {Mazziotta}, {McEnery}, {Meurer},
  {Michelson}, {Mitthumsiri}, {Mizuno}, {Monte}, {Monzani}, {Morselli},
  {Moskalenko}, {Murgia}, {Nolan}, {Norris}, {Nuss}, {Ohsugi}, {Okumura},
  {Omodei}, {Orlando}, {Ormes}, {Paneque}, {Pelassa}, {Pepe}, {Pesce-Rollins},
  {Piron}, {Porter}, {Rain{\`o}}, {Rando}, {Razzano}, {Reimer}, {Reimer},
  {Reposeur}, {Rodriguez}, {Ryde}, {Sadrozinski}, {Sanchez}, {Sander}, {Saz
  Parkinson}, {Sgr{\`o}}, {Siskind}, {Smith}, {Spandre}, {Spinelli}, {Starck},
  {Strickman}, {Strong}, {Suson}, {Takahashi}, {Tanaka}, {Thayer}, {Thayer},
  {Thompson}, {Tibaldo}, {Torres}, {Tosti}, {Tramacere}, {Uchiyama}, {Usher},
  {Vasileiou}, {Vilchez}, {Vitale}, {Waite}, {Wang}, {Winer}, {Wood}, {Ylinen},
  {Ziegler}, \& {Fermi/LAT Collaboration}}]{fermi2010}
{Abdo}, A.~A., {Ackermann}, M., {Ajello}, M., {et~al.} 2010, \apj, 710, 133

\bibitem[{{Alves} {et~al.}(2007){Alves}, {Lombardi}, \& {Lada}}]{alves07}
{Alves}, J., {Lombardi}, M., \& {Lada}, C.~J. 2007, \aap, 462, L17

\bibitem[{{Bains} {et~al.}(2006){Bains}, {Wong}, {Cunningham}, {Sparks},
  {Brisbin}, {Calisse}, {Dempsey}, {Deragopian}, {Ellingsen}, {Fulton},
  {Herpin}, {Jones}, {Kouba}, {Kramer}, {Ladd}, {Longmore}, {McEvoy}, {Maller},
  {Minier}, {Mookerjea}, {Phillips}, {Purcell}, {Walsh}, {Voronkov}, \&
  {Burton}}]{bains06}
{Bains}, I., {Wong}, T., {Cunningham}, M., {et~al.} 2006, \mnras, 367, 1609

\bibitem[{{Barnes} {et~al.}(2013){Barnes}, {Muller}, {Inderm\''{u}hle},
  {O'Dougherty}, {Lowe}, {Cunningham}, {Hernandez}, \& {Fuller}}]{barnes13}
{Barnes}, P., {Muller}, E., {Inderm\''{u}hle}, B., {et~al.} 2013, \apj

\bibitem[{{Brandner} {et~al.}(2008){Brandner}, {Clark}, {Stolte}, {Waters},
  {Negueruela}, \& {Goodwin}}]{brandner08}
{Brandner}, W., {Clark}, J.~S., {Stolte}, A., {et~al.} 2008, \aap, 478, 137

\bibitem[{{Bronfman} {et~al.}(1989){Bronfman}, {Alvarez}, {Cohen}, \&
  {Thaddeus}}]{bronfman1989}
{Bronfman}, L., {Alvarez}, H., {Cohen}, R.~S., \& {Thaddeus}, P. 1989, \apjs,
  71, 481

\bibitem[{{Carey} {et~al.}(2009){Carey}, {Noriega-Crespo}, {Mizuno}, {Shenoy},
  {Paladini}, {Kraemer}, {Price}, {Flagey}, {Ryan}, {Ingalls}, {Kuchar},
  {Pinheiro Gon{\c c}alves}, {Indebetouw}, {Billot}, {Marleau}, {Padgett},
  {Rebull}, {Bressert}, {Ali}, {Molinari}, {Martin}, {Berriman}, {Boulanger},
  {Latter}, {Miville-Deschenes}, {Shipman}, \& {Testi}}]{carey09}
{Carey}, S.~J., {Noriega-Crespo}, A., {Mizuno}, D.~R., {et~al.} 2009, \pasp,
  121, 76

\bibitem[{{Caswell} \& {Haynes}(1975)}]{caswell75}
{Caswell}, J.~L. \& {Haynes}, R.~F. 1975, \mnras, 173, 649

\bibitem[{{Caswell} {et~al.}(1980){Caswell}, {Haynes}, \& {Goss}}]{caswell1980}
{Caswell}, J.~L., {Haynes}, R.~F., \& {Goss}, W.~M. 1980, Australian Journal of
  Physics, 33, 639

\bibitem[{{Chabrier} {et~al.}(2014){Chabrier}, {Hennebelle}, \&
  {Charlot}}]{chabrier14}
{Chabrier}, G., {Hennebelle}, P., \& {Charlot}, S. 2014, \apj, 796, 75

\bibitem[{{Dahmen} {et~al.}(1998){Dahmen}, {Huttemeister}, {Wilson}, \&
  {Mauersberger}}]{dahmen98}
{Dahmen}, G., {Huttemeister}, S., {Wilson}, T.~L., \& {Mauersberger}, R. 1998,
  \aap, 331, 959

\bibitem[{{Dame} {et~al.}(2001){Dame}, {Hartmann}, \& {Thaddeus}}]{dame2001}
{Dame}, T.~M., {Hartmann}, D., \& {Thaddeus}, P. 2001, \apj, 547, 792

\bibitem[{{Dibai}(1958)}]{dibai58}
{Dibai}, E.~A. 1958, \sovast, 2, 226

\bibitem[{{Dobashi} {et~al.}(2014){Dobashi}, {Matsumoto}, {Shimoikura},
  {Saito}, {Akisato}, {Ohashi}, \& {Nakagomi}}]{dobashi14}
{Dobashi}, K., {Matsumoto}, T., {Shimoikura}, T., {et~al.} 2014, \apj, 797, 58

\bibitem[{{Dobbs} \& {Pringle}(2013)}]{dobbs13}
{Dobbs}, C.~L. \& {Pringle}, J.~E. 2013, \mnras, 432, 653

\bibitem[{{Dutra} {et~al.}(2003){Dutra}, {Bica}, {Soares}, \&
  {Barbuy}}]{dutra03}
{Dutra}, C.~M., {Bica}, E., {Soares}, J., \& {Barbuy}, B. 2003, \aap, 400, 533

\bibitem[{{Eden} {et~al.}(2013){Eden}, {Moore}, {Morgan}, {Thompson}, \&
  {Urquhart}}]{eden13}
{Eden}, D.~J., {Moore}, T.~J.~T., {Morgan}, L.~K., {Thompson}, M.~A., \&
  {Urquhart}, J.~S. 2013, \mnras, 431, 1587

\bibitem[{{Eden} {et~al.}(2012){Eden}, {Moore}, {Plume}, \& {Morgan}}]{eden12}
{Eden}, D.~J., {Moore}, T.~J.~T., {Plume}, R., \& {Morgan}, L.~K. 2012, \mnras,
  422, 3178

\bibitem[{{Fujiyoshi} {et~al.}(2006){Fujiyoshi}, {Smith}, {Caswell}, {Moore},
  {Lumsden}, {Aitken}, \& {Roche}}]{fujiyoshi06}
{Fujiyoshi}, T., {Smith}, C.~H., {Caswell}, J.~L., {et~al.} 2006, \mnras, 368,
  1843

\bibitem[{{Fujiyoshi} {et~al.}(2005){Fujiyoshi}, {Smith}, {Moore}, {Lumsden},
  {Aitken}, \& {Roche}}]{fujiyoshi05}
{Fujiyoshi}, T., {Smith}, C.~H., {Moore}, T.~J.~T., {et~al.} 2005, \mnras, 356,
  801

\bibitem[{{Georgelin} \& {Georgelin}(1976)}]{georgelin1976}
{Georgelin}, Y.~M. \& {Georgelin}, Y.~P. 1976, \aap, 49, 57

\bibitem[{{Grave} {et~al.}(2014){Grave}, {Kumar}, {Ojha}, {Teixeira}, \&
  {Pace}}]{grave14}
{Grave}, J.~M.~C., {Kumar}, M.~S.~N., {Ojha}, D.~K., {Teixeira}, G.~D.~C., \&
  {Pace}, G. 2014, \aap, 563, A123

\bibitem[{{Green}(2009)}]{green09}
{Green}, D.~A. 2009, Bulletin of the Astronomical Society of India, 37, 45

\bibitem[{{Griffin} {et~al.}(2010){Griffin}, {Abergel}, {Abreu}, {Ade},
  {Andr{\'e}}, {Augueres}, {Babbedge}, {Bae}, {Baillie}, {Baluteau}, {Barlow},
  {Bendo}, {Benielli}, {Bock}, {Bonhomme}, {Brisbin}, {Brockley-Blatt},
  {Caldwell}, {Cara}, {Castro-Rodriguez}, {Cerulli}, {Chanial}, {Chen},
  {Clark}, {Clements}, {Clerc}, {Coker}, {Communal}, {Conversi}, {Cox},
  {Crumb}, {Cunningham}, {Daly}, {Davis}, {de Antoni}, {Delderfield}, {Devin},
  {di Giorgio}, {Didschuns}, {Dohlen}, {Donati}, {Dowell}, {Dowell}, {Duband},
  {Dumaye}, {Emery}, {Ferlet}, {Ferrand}, {Fontignie}, {Fox}, {Franceschini},
  {Frerking}, {Fulton}, {Garcia}, {Gastaud}, {Gear}, {Glenn}, {Goizel},
  {Griffin}, {Grundy}, {Guest}, {Guillemet}, {Hargrave}, {Harwit}, {Hastings},
  {Hatziminaoglou}, {Herman}, {Hinde}, {Hristov}, {Huang}, {Imhof}, {Isaak},
  {Israelsson}, {Ivison}, {Jennings}, {Kiernan}, {King}, {Lange}, {Latter},
  {Laurent}, {Laurent}, {Leeks}, {Lellouch}, {Levenson}, {Li}, {Li},
  {Lilienthal}, {Lim}, {Liu}, {Lu}, {Madden}, {Mainetti}, {Marliani}, {McKay},
  {Mercier}, {Molinari}, {Morris}, {Moseley}, {Mulder}, {Mur}, {Naylor},
  {Nguyen}, {O'Halloran}, {Oliver}, {Olofsson}, {Olofsson}, {Orfei}, {Page},
  {Pain}, {Panuzzo}, {Papageorgiou}, {Parks}, {Parr-Burman}, {Pearce},
  {Pearson}, {P{\'e}rez-Fournon}, {Pinsard}, {Pisano}, {Podosek}, {Pohlen},
  {Polehampton}, {Pouliquen}, {Rigopoulou}, {Rizzo}, {Roseboom}, {Roussel},
  {Rowan-Robinson}, {Rownd}, {Saraceno}, {Sauvage}, {Savage}, {Savini},
  {Sawyer}, {Scharmberg}, {Schmitt}, {Schneider}, {Schulz}, {Schwartz},
  {Shafer}, {Shupe}, {Sibthorpe}, {Sidher}, {Smith}, {Smith}, {Smith},
  {Spencer}, {Stobie}, {Sudiwala}, {Sukhatme}, {Surace}, {Stevens}, {Swinyard},
  {Trichas}, {Tourette}, {Triou}, {Tseng}, {Tucker}, {Turner}, {Vaccari},
  {Valtchanov}, {Vigroux}, {Virique}, {Voellmer}, {Walker}, {Ward}, {Waskett},
  {Weilert}, {Wesson}, {White}, {Whitehouse}, {Wilson}, {Winter}, {Woodcraft},
  {Wright}, {Xu}, {Zavagno}, {Zemcov}, {Zhang}, \& {Zonca}}]{griffin10}
{Griffin}, M.~J., {Abergel}, A., {Abreu}, A., {et~al.} 2010, \aap, 518, L3+

\bibitem[{{Gritschneder} {et~al.}(2010){Gritschneder}, {Burkert}, {Naab}, \&
  {Walch}}]{gritschneder10}
{Gritschneder}, M., {Burkert}, A., {Naab}, T., \& {Walch}, S. 2010, \apj, 723,
  971

\bibitem[{{Heiderman} {et~al.}(2010){Heiderman}, {Evans}, {Allen}, {Huard}, \&
  {Heyer}}]{heiderman10}
{Heiderman}, A., {Evans}, II, N.~J., {Allen}, L.~E., {Huard}, T., \& {Heyer},
  M. 2010, \apj, 723, 1019

\bibitem[{{Hennebelle} {et~al.}(2001){Hennebelle}, {P{\'e}rault}, {Teyssier},
  \& {Ganesh}}]{hennebelle01}
{Hennebelle}, P., {P{\'e}rault}, M., {Teyssier}, D., \& {Ganesh}, S. 2001,
  \aap, 365, 598

\bibitem[{{Immer} {et~al.}(2012){Immer}, {Menten}, {Schuller}, \&
  {Lis}}]{immer12}
{Immer}, K., {Menten}, K.~M., {Schuller}, F., \& {Lis}, D.~C. 2012, \aap, 548,
  A120

\bibitem[{{Inutsuka} \& {Miyama}(1992)}]{inutsuka92}
{Inutsuka}, S. \& {Miyama}, S.~M. 1992, \apj, 388, 392

\bibitem[{{Jackson} {et~al.}(2006){Jackson}, {Rathborne}, {Shah}, {Simon},
  {Bania}, {Clemens}, {Chambers}, {Johnson}, {Dormody}, {Lavoie}, \&
  {Heyer}}]{jackson06}
{Jackson}, J.~M., {Rathborne}, J.~M., {Shah}, R.~Y., {et~al.} 2006, \apjs, 163,
  145

\bibitem[{{Jones} \& {Dickey}(2012)}]{jones12a}
{Jones}, C. \& {Dickey}, J.~M. 2012, \apj, 753, 62

\bibitem[{{Jones} {et~al.}(2012){Jones}, {Burton}, {Cunningham},
  {Requena-Torres}, {Menten}, {Schilke}, {Belloche}, {Leurini},
  {Mart{\'{\i}}n-Pintado}, {Ott}, \& {Walsh}}]{jones12}
{Jones}, P.~A., {Burton}, M.~G., {Cunningham}, M.~R., {et~al.} 2012, \mnras,
  419, 2961

\bibitem[{{Karnik} {et~al.}(2001){Karnik}, {Ghosh}, {Rengarajan}, \&
  {Verma}}]{karnick2001}
{Karnik}, A.~D., {Ghosh}, S.~K., {Rengarajan}, T.~N., \& {Verma}, R.~P. 2001,
  \mnras, 326, 293

\bibitem[{{Kennicutt}(1998)}]{kennicutt98}
{Kennicutt}, Jr., R.~C. 1998, \apj, 498, 541

\bibitem[{{Kim} {et~al.}(2006){Kim}, {Figer}, {Kudritzki}, \&
  {Najarro}}]{kim06}
{Kim}, S.~S., {Figer}, D.~F., {Kudritzki}, R.~P., \& {Najarro}, F. 2006, \apjl,
  653, L113

\bibitem[{{Kroupa}(2001)}]{kroupa01}
{Kroupa}, P. 2001, \mnras, 322, 231

\bibitem[{{Kruijssen} {et~al.}(2014){Kruijssen}, {Longmore}, {Elmegreen},
  {Murray}, {Bally}, {Testi}, \& {Kennicutt}}]{kruijssen14}
{Kruijssen}, J.~M.~D., {Longmore}, S.~N., {Elmegreen}, B.~G., {et~al.} 2014,
  \mnras, 440, 3370

\bibitem[{{Krumholz} \& {Matzner}(2009)}]{krumholz09b}
{Krumholz}, M.~R. \& {Matzner}, C.~D. 2009, \apj, 703, 1352

\bibitem[{{Kuchar} \& {Clark}(1997)}]{kc97c}
{Kuchar}, T.~A. \& {Clark}, F.~O. 1997, \apj, 488, 224

\bibitem[{{Kumar}(2013)}]{kumar13}
{Kumar}, M.~S.~N. 2013, \aap, 558, A119

\bibitem[{{Lo} {et~al.}(2007){Lo}, {Cunningham}, {Bains}, {Burton}, \&
  {Garay}}]{lo07}
{Lo}, N., {Cunningham}, M., {Bains}, I., {Burton}, M.~G., \& {Garay}, G. 2007,
  \mnras, 381, L30

\bibitem[{{Lo} {et~al.}(2009){Lo}, {Cunningham}, {Jones}, {Bains}, {Burton},
  {Wong}, {Muller}, {Kramer}, {Ossenkopf}, {Henkel}, {Deragopian}, {Donnelly},
  \& {Ladd}}]{lo09}
{Lo}, N., {Cunningham}, M.~R., {Jones}, P.~A., {et~al.} 2009, \mnras, 395, 1021

\bibitem[{{Lo} {et~al.}(2011){Lo}, {Redman}, {Jones}, {Cunningham}, {Chhetri},
  {Bains}, \& {Burton}}]{lo11}
{Lo}, N., {Redman}, M.~P., {Jones}, P.~A., {et~al.} 2011, \mnras, 415, 525

\bibitem[{{Lockman}(1979)}]{lockman79}
{Lockman}, F.~J. 1979, \apj, 232, 761

\bibitem[{{Louvet} {et~al.}(2014){Louvet}, {Motte}, {Hennebelle}, {Maury},
  {Bonnell}, {Bontemps}, {Gusdorf}, {Hill}, {Gueth}, {Peretto},
  {Duarte-Cabral}, {Stephan}, {Schilke}, {Csengeri}, {Nguyen Luong}, \&
  {Lis}}]{louvet14}
{Louvet}, F., {Motte}, F., {Hennebelle}, P., {et~al.} 2014, \aap, 570, A15

\bibitem[{{Lowe} {et~al.}(2014){Lowe}, {Cunningham}, {Urquhart}, {Marshall},
  {Horiuchi}, {Lo}, {Walsh}, {Jordan}, {Jones}, \& {Hill}}]{lowe14}
{Lowe}, V., {Cunningham}, M.~R., {Urquhart}, J.~S., {et~al.} 2014, \mnras, 441,
  256

\bibitem[{{Lynga}(1964)}]{lynga64}
{Lynga}, G. 1964, Meddelanden fran Lunds Astronomiska Observatorium Serie II,
  141, 1

\bibitem[{{Martins} {et~al.}(2008){Martins}, {Hillier}, {Paumard},
  {Eisenhauer}, {Ott}, \& {Genzel}}]{martins08}
{Martins}, F., {Hillier}, D.~J., {Paumard}, T., {et~al.} 2008, \aap, 478, 219

\bibitem[{{Martins} {et~al.}(2005){Martins}, {Schaerer}, \&
  {Hillier}}]{martins05}
{Martins}, F., {Schaerer}, D., \& {Hillier}, D.~J. 2005, \aap, 436, 1049

\bibitem[{{McClure-Griffiths} {et~al.}(2005){McClure-Griffiths}, {Dickey},
  {Gaensler}, {Green}, {Haverkorn}, \& {Strasser}}]{mcclure05}
{McClure-Griffiths}, N.~M., {Dickey}, J.~M., {Gaensler}, B.~M., {et~al.} 2005,
  \apjs, 158, 178

\bibitem[{{McClure-Griffiths} {et~al.}(2001){McClure-Griffiths}, {Green},
  {Dickey}, {Gaensler}, {Haynes}, \& {Wieringa}}]{mcclure01}
{McClure-Griffiths}, N.~M., {Green}, A.~J., {Dickey}, J.~M., {et~al.} 2001,
  \apj, 551, 394

\bibitem[{{Mel'Nik} \& {Efremov}(1995)}]{melnik95}
{Mel'Nik}, A.~M. \& {Efremov}, Y.~N. 1995, Astronomy Letters, 21, 10

\bibitem[{{Mercer} {et~al.}(2005){Mercer}, {Clemens}, {Meade}, {Babler},
  {Indebetouw}, {Whitney}, {Watson}, {Wolfire}, {Wolff}, {Bania}, {Benjamin},
  {Cohen}, {Dickey}, {Jackson}, {Kobulnicky}, {Mathis}, {Stauffer}, {Stolovy},
  {Uzpen}, \& {Churchwell}}]{mcm05b}
{Mercer}, E.~P., {Clemens}, D.~P., {Meade}, M.~R., {et~al.} 2005, \apj, 635,
  560

\bibitem[{{Merello} {et~al.}(2013{\natexlab{a}}){Merello}, {Bronfman}, {Garay},
  {Lo}, {Evans}, {Nyman}, {Cort{\'e}s}, \& {Cunningham}}]{merello2013alma}
{Merello}, M., {Bronfman}, L., {Garay}, G., {et~al.} 2013{\natexlab{a}}, \apjl,
  774, L7

\bibitem[{{Merello} {et~al.}(2013{\natexlab{b}}){Merello}, {Bronfman}, {Garay},
  {Nyman}, {Evans}, \& {Walmsley}}]{merello2013}
{Merello}, M., {Bronfman}, L., {Garay}, G., {et~al.} 2013{\natexlab{b}}, \apj,
  774, 38

\bibitem[{{Mezger} \& {Henderson}(1967)}]{mezger67}
{Mezger}, P.~G. \& {Henderson}, A.~P. 1967, \apj, 147, 471

\bibitem[{{Miyazaki} \& {Tsuboi}(2000)}]{miyazaki00}
{Miyazaki}, A. \& {Tsuboi}, M. 2000, \apj, 536, 357

\bibitem[{{Mois{\'e}s} {et~al.}(2011){Mois{\'e}s}, {Damineli}, {Figuer{\^e}do},
  {Blum}, {Conti}, \& {Barbosa}}]{moises11}
{Mois{\'e}s}, A.~P., {Damineli}, A., {Figuer{\^e}do}, E., {et~al.} 2011,
  \mnras, 411, 705

\bibitem[{{Molinari} {et~al.}(2010){Molinari}, {Swinyard}, {Bally}, {Barlow},
  {Bernard}, {Martin}, {Moore}, {Noriega-Crespo}, {Plume}, {Testi}, {Zavagno},
  {Abergel}, {Ali}, {Anderson}, {Andr{\'e}}, {Baluteau}, {Battersby},
  {Beltr{\'a}n}, {Benedettini}, {Billot}, {Blommaert}, {Bontemps}, {Boulanger},
  {Brand}, {Brunt}, {Burton}, {Calzoletti}, {Carey}, {Caselli}, {Cesaroni},
  {Cernicharo}, {Chakrabarti}, {Chrysostomou}, {Cohen}, {Compiegne}, {de
  Bernardis}, {de Gasperis}, {di Giorgio}, {Elia}, {Faustini}, {Flagey},
  {Fukui}, {Fuller}, {Ganga}, {Garcia-Lario}, {Glenn}, {Goldsmith}, {Griffin},
  {Hoare}, {Huang}, {Ikhenaode}, {Joblin}, {Joncas}, {Juvela}, {Kirk},
  {Lagache}, {Li}, {Lim}, {Lord}, {Marengo}, {Marshall}, {Masi}, {Massi},
  {Matsuura}, {Minier}, {Miville-Desch{\^e}nes}, {Montier}, {Morgan}, {Motte},
  {Mottram}, {M{\"u}ller}, {Natoli}, {Neves}, {Olmi}, {Paladini}, {Paradis},
  {Parsons}, {Peretto}, {Pestalozzi}, {Pezzuto}, {Piacentini}, {Piazzo},
  {Polychroni}, {Pomar{\`e}s}, {Popescu}, {Reach}, {Ristorcelli}, {Robitaille},
  {Robitaille}, {Rod{\'o}n}, {Roy}, {Royer}, {Russeil}, {Saraceno}, {Sauvage},
  {Schilke}, {Schisano}, {Schneider}, {Schuller}, {Schulz}, {Sibthorpe},
  {Smith}, {Smith}, {Spinoglio}, {Stamatellos}, {Strafella}, {Stringfellow},
  {Sturm}, {Taylor}, {Thompson}, {Traficante}, {Tuffs}, {Umana}, {Valenziano},
  {Vavrek}, {Veneziani}, {Viti}, {Waelkens}, {Ward-Thompson}, {White},
  {Wilcock}, {Wyrowski}, {Yorke}, \& {Zhang}}]{molinari10}
{Molinari}, S., {Swinyard}, B., {Bally}, J., {et~al.} 2010, \aap, 518, L100+

\bibitem[{{Mookerjea} {et~al.}(2004){Mookerjea}, {Kramer}, {Nielbock}, \&
  {Nyman}}]{mooker04}
{Mookerjea}, B., {Kramer}, C., {Nielbock}, M., \& {Nyman}, L.-{\AA}. 2004,
  \aap, 426, 119

\bibitem[{{Motte} {et~al.}(2007){Motte}, {Bontemps}, {Schilke}, {Schneider},
  {Menten}, \& {Brogui{\`e}re}}]{motte07}
{Motte}, F., {Bontemps}, S., {Schilke}, P., {et~al.} 2007, \aap, 476, 1243

\bibitem[{{Motte} {et~al.}(2014){Motte}, {Nguy{\^e}n Luong}, {Schneider},
  {Heitsch}, {Glover}, {Carlhoff}, {Hill}, {Bontemps}, {Schilke}, {Louvet},
  {Hennemann}, {Didelon}, \& {Beuther}}]{motte14}
{Motte}, F., {Nguy{\^e}n Luong}, Q., {Schneider}, N., {et~al.} 2014, \aap, 571,
  A32

\bibitem[{{Murray}(2011)}]{murray11}
{Murray}, N. 2011, \apj, 729, 133

\bibitem[{{Myers}(2014)}]{myers14}
{Myers}, P.~C. 2014, \apj, 781, 33

\bibitem[{{Nguyen Luong} {et~al.}(2011{\natexlab{a}}){Nguyen Luong}, {Motte},
  {Hennemann}, {Hill}, {Rygl}, {Schneider}, {Bontemps}, {Men'shchikov},
  {Andr{\'e}}, {Peretto}, {Anderson}, {Arzoumanian}, {Deharveng}, {Didelon},
  {di Francesco}, {Griffin}, {Kirk}, {K{\"o}nyves}, {Martin}, {Maury},
  {Minier}, {Molinari}, {Pestalozzi}, {Pezzuto}, {Reid}, {Roussel}, {Sauvage},
  {Schuller}, {Testi}, {Ward-Thompson}, {White}, \& {Zavagno}}]{nguyenluong11b}
{Nguyen Luong}, Q., {Motte}, F., {Hennemann}, M., {et~al.} 2011{\natexlab{a}},
  \aap, 535, A76

\bibitem[{{Nguyen Luong} {et~al.}(2011{\natexlab{b}}){Nguyen Luong}, {Motte},
  {Schuller}, {Schneider}, {Bontemps}, {Schilke}, {Menten}, {Heitsch},
  {Wyrowski}, {Carlhoff}, {Bronfman}, \& {Henning}}]{nguyenluong11}
{Nguyen Luong}, Q., {Motte}, F., {Schuller}, F., {et~al.} 2011{\natexlab{b}},
  \aap, 529, A41+

\bibitem[{{Ostriker}(1964)}]{ostriker64}
{Ostriker}, J. 1964, \apj, 140, 1056

\bibitem[{{Ozernoi}(1964)}]{ozernoi64}
{Ozernoi}, L.~M. 1964, \sovast, 8, 137

\bibitem[{{Parmentier} \& {Pfalzner}(2013)}]{parmentier13}
{Parmentier}, G. \& {Pfalzner}, S. 2013, \aap, 549, A132

\bibitem[{{Peeters} {et~al.}(2004){Peeters}, {Spoon}, \& {Tielens}}]{peeters04}
{Peeters}, E., {Spoon}, H.~W.~W., \& {Tielens}, A.~G.~G.~M. 2004, \apj, 613,
  986

\bibitem[{{Peretto} {et~al.}(2010){Peretto}, {Fuller}, {Plume}, {Anderson},
  {Bally}, {Battersby}, {Beltran}, {Bernard}, {Calzoletti}, {Digiorgio},
  {Faustini}, {Kirk}, {Lenfestey}, {Marshall}, {Martin}, {Molinari}, {Montier},
  {Motte}, {Ristorcelli}, {Rod{\'o}n}, {Smith}, {Traficante}, {Veneziani},
  {Ward-Thompson}, \& {Wilcock}}]{peretto10c}
{Peretto}, N., {Fuller}, G.~A., {Plume}, R., {et~al.} 2010, \aap, 518, L98+

\bibitem[{{Poglitsch} {et~al.}(2010){Poglitsch}, {Waelkens}, {Geis},
  {Feuchtgruber}, {Vandenbussche}, {Rodriguez}, {Krause}, {Renotte}, {van
  Hoof}, {Saraceno}, {Cepa}, {Kerschbaum}, {Agn{\`e}se}, {Ali}, {Altieri},
  {Andreani}, {Augueres}, {Balog}, {Barl}, {Bauer}, {Belbachir}, {Benedettini},
  {Billot}, {Boulade}, {Bischof}, {Blommaert}, {Callut}, {Cara}, {Cerulli},
  {Cesarsky}, {Contursi}, {Creten}, {De Meester}, {Doublier}, {Doumayrou},
  {Duband}, {Exter}, {Genzel}, {Gillis}, {Gr{\"o}zinger}, {Henning},
  {Herreros}, {Huygen}, {Inguscio}, {Jakob}, {Jamar}, {Jean}, {de Jong},
  {Katterloher}, {Kiss}, {Klaas}, {Lemke}, {Lutz}, {Madden}, {Marquet},
  {Martignac}, {Mazy}, {Merken}, {Montfort}, {Morbidelli}, {M{\"u}ller},
  {Nielbock}, {Okumura}, {Orfei}, {Ottensamer}, {Pezzuto}, {Popesso},
  {Putzeys}, {Regibo}, {Reveret}, {Royer}, {Sauvage}, {Schreiber}, {Stegmaier},
  {Schmitt}, {Schubert}, {Sturm}, {Thiel}, {Tofani}, {Vavrek}, {Wetzstein},
  {Wieprecht}, \& {Wiezorrek}}]{poglitsch10}
{Poglitsch}, A., {Waelkens}, C., {Geis}, N., {et~al.} 2010, \aap, 518, L2+

\bibitem[{{Portegies Zwart} {et~al.}(2010){Portegies Zwart}, {McMillan}, \&
  {Gieles}}]{portegies-zwart10}
{Portegies Zwart}, S.~F., {McMillan}, S.~L.~W., \& {Gieles}, M. 2010, \araa,
  48, 431

\bibitem[{{Rathborne} {et~al.}(2006){Rathborne}, {Jackson}, \&
  {Simon}}]{rathborne06}
{Rathborne}, J.~M., {Jackson}, J.~M., \& {Simon}, R. 2006, \apj, 641, 389

\bibitem[{{Reid} {et~al.}(2009){Reid}, {Menten}, {Zheng}, {Brunthaler},
  {Moscadelli}, {Xu}, {Zhang}, {Sato}, {Honma}, {Hirota}, {Hachisuka}, {Choi},
  {Moellenbrock}, \& {Bartkiewicz}}]{reid2009}
{Reid}, M.~J., {Menten}, K.~M., {Zheng}, X.~W., {et~al.} 2009, \apj, 700, 137

\bibitem[{{Robitaille} {et~al.}(2008){Robitaille}, {Meade}, {Babler},
  {Whitney}, {Johnston}, {Indebetouw}, {Cohen}, {Povich}, {Sewilo}, {Benjamin},
  \& {Churchwell}}]{robitaille08}
{Robitaille}, T.~P., {Meade}, M.~R., {Babler}, B.~L., {et~al.} 2008, \aj, 136,
  2413

\bibitem[{{Rodgers} {et~al.}(1960){Rodgers}, {Campbell}, \&
  {Whiteoak}}]{rodgers1960}
{Rodgers}, A.~W., {Campbell}, C.~T., \& {Whiteoak}, J.~B. 1960, \mnras, 121,
  103

\bibitem[{{Rodriguez-Fernandez} \& {Combes}(2008)}]{Rodriguez-Fernandez08}
{Rodriguez-Fernandez}, N.~J. \& {Combes}, F. 2008, \aap, 489, 115

\bibitem[{{Roman-Duval} {et~al.}(2009){Roman-Duval}, {Jackson}, {Heyer},
  {Johnson}, {Rathborne}, {Shah}, \& {Simon}}]{romanduval09}
{Roman-Duval}, J., {Jackson}, J.~M., {Heyer}, M., {et~al.} 2009, \apj, 699,
  1153

\bibitem[{{Roussel}(2013)}]{roussel13}
{Roussel}, H. 2013, \pasp, 125, 1126

\bibitem[{{Russeil} {et~al.}(2005){Russeil}, {Adami}, {Amram}, {Le Coarer},
  {Georgelin}, {Marcelin}, \& {Parker}}]{russeil05}
{Russeil}, D., {Adami}, C., {Amram}, P., {et~al.} 2005, \aap, 429, 497

\bibitem[{{Russeil} {et~al.}(2012){Russeil}, {Zavagno}, {Adami}, {Anderson},
  {Bontemps}, {Motte}, {Rodon}, {Schneider}, {Ilmane}, \& {Murphy}}]{russeil12}
{Russeil}, D., {Zavagno}, A., {Adami}, C., {et~al.} 2012, \aap, 538, A142

\bibitem[{{Sakamoto} {et~al.}(1997){Sakamoto}, {Hasegawa}, {Handa}, {Hayashi},
  \& {Oka}}]{sakamoto1997}
{Sakamoto}, S., {Hasegawa}, T., {Handa}, T., {Hayashi}, M., \& {Oka}, T. 1997,
  \apj, 486, 276

\bibitem[{{Salpeter}(1955)}]{salpeter55}
{Salpeter}, E.~E. 1955, \apj, 121, 161

\bibitem[{{Schneider} {et~al.}(2006){Schneider}, {Bontemps}, {Simon}, {Jakob},
  {Motte}, {Miller}, {Kramer}, \& {Stutzki}}]{schneider06}
{Schneider}, N., {Bontemps}, S., {Simon}, R., {et~al.} 2006, \aap, 458, 855

\bibitem[{{Simon} {et~al.}(2006){Simon}, {Rathborne}, {Shah}, {Jackson}, \&
  {Chambers}}]{simon06b}
{Simon}, R., {Rathborne}, J.~M., {Shah}, R.~Y., {Jackson}, J.~M., \&
  {Chambers}, E.~T. 2006, \apj, 653, 1325

\bibitem[{{Stolte} {et~al.}(2005){Stolte}, {Brandner}, {Grebel}, {Lenzen}, \&
  {Lagrange}}]{stolte05}
{Stolte}, A., {Brandner}, W., {Grebel}, E.~K., {Lenzen}, R., \& {Lagrange},
  A.-M. 2005, \apjl, 628, L113

\bibitem[{{Stolte} {et~al.}(2002){Stolte}, {Grebel}, {Brandner}, \&
  {Figer}}]{stolte02}
{Stolte}, A., {Grebel}, E.~K., {Brandner}, W., \& {Figer}, D.~F. 2002, \aap,
  394, 459

\bibitem[{{Thompson}(1984)}]{thompson1984}
{Thompson}, R.~I. 1984, \apj, 283, 165

\bibitem[{{Tremblin} {et~al.}(2013){Tremblin}, {Minier}, {Schneider}, {Audit},
  {Hill}, {Didelon}, {Peretto}, {Arzoumanian}, {Motte}, {Zavagno}, {Bontemps},
  {Anderson}, {Andr{\'e}}, {Bernard}, {Csengeri}, {Di Francesco}, {Elia},
  {Hennemann}, {K{\"o}nyves}, {Marston}, {Nguyen Luong}, {Rivera-Ingraham},
  {Roussel}, {Sousbie}, {Spinoglio}, {White}, \& {Williams}}]{tremblin13}
{Tremblin}, P., {Minier}, V., {Schneider}, N., {et~al.} 2013, \aap, 560, A19

\bibitem[{{Urquhart} {et~al.}(2007){Urquhart}, {Busfield}, {Hoare}, {Lumsden},
  {Clarke}, {Moore}, {Mottram}, \& {Oudmaijer}}]{urqhart07}
{Urquhart}, J.~S., {Busfield}, A.~L., {Hoare}, M.~G., {et~al.} 2007, \aap, 461,
  11

\bibitem[{{Walsh} {et~al.}(1997){Walsh}, {Hyland}, {Robinson}, \&
  {Burton}}]{walsh97}
{Walsh}, A.~J., {Hyland}, A.~R., {Robinson}, G., \& {Burton}, M.~G. 1997,
  \mnras, 291, 261

\bibitem[{{Wilson} {et~al.}(2012){Wilson}, {Casassus}, \& {Keating}}]{wilson12}
{Wilson}, T.~L., {Casassus}, S., \& {Keating}, K.~M. 2012, \apj, 744, 161

\bibitem[{{Wong} {et~al.}(2008){Wong}, {Ladd}, {Brisbin}, {Burton}, {Bains},
  {Cunningham}, {Lo}, {Jones}, {Thomas}, {Longmore}, {Vigan}, {Mookerjea},
  {Kramer}, {Fukui}, \& {Kawamura}}]{wong08}
{Wong}, T., {Ladd}, E.~F., {Brisbin}, D., {et~al.} 2008, \mnras, 386, 1069

\end{thebibliography}

\appendix

\section{21 cm continuum sources}\label{AppendixA}

In Fig.~\ref{21cm_box}, we selected 31 rectangular subregions in the
21~cm continuum data in which there is significant emission.  These
are the ``boxes'' whose details are captured in Table~\ref{box_table}.

\begin{table*}[!tbhp]
\scriptsize
\centering
\caption{Details of boxes and subregions as selected from the 21 cm continuum map}
\setlength{\tabcolsep}{0.07cm}
\renewcommand{\arraystretch}{0.75}
\begin{threeparttable}
\begin{tabular}{c|cccccccccccccc}
\hline
\hline
\noalign{\vskip 2pt}
No & $\ell$ & $b$ & Type & \vlsr & Cloud & Ang. size & Area & $S_\nu$(21\,cm) & SFR density & Mass & $\Sigma_{\rm gas}$ & $\log N_{\rm Ly\alpha}$ & \# O7V & $M_{tot}$ O7V \\

& (\degr) & (\degr) & & (\kms) & & (\arcmin) & (pc$^{2}$) & (Jy) &  (\sfrdens) & (M$_\odot$) & (\mpcs) & (s$^{-1}$) & & (M$_\odot$) \\
\noalign{\vskip 2pt}
\hline
\noalign{\vskip 2pt}

1 & 334.7 & 0 & C\hii & -60 to -30 & RCW~106 & 2.6 & - & 0.48 & 0.73 & - & - & 47.68 &0 & 0 \\

2 & 334.7 & -0.6 & C\hii & -60 to -30 & RCW~106 & 2.9 & - & 0.96 & 1.3 & - & - & 48 & 0 & 0 \\

3 & 334.6 & -0.1 & C\hii & -60 to -30 & RCW~106 & 5.5 & - & 1.8 & 0.78 & - & - & 48.31 & 0 & 0 \\

4 & 334.6 & 0.4 & C\hii & -60 to -30 & RCW~106 & 1.6 & - & 0.12 & 0.53 & - & - & 47.21 &0 & 0 \\

5 & 334.5 & 0.8 & C\hii & -60 to -30 & RCW~106 & 3.7 & - & 0.29 & 0.14 & - & - & 47.76 & 0 & 0 \\

6 & 334.2 & 0.1 & SNR & - & - & 27.7 & - & 6.4 & 0.24 & - & - & 48.88 &- & - \\

7 & 333.6 & -0.2 & G\hii & -97 to -78 & \gthree & 31.5 & 2109.35 & - & - & 1.7
$\times10^{5}$ & 80.95 & - &- & - \\

& & & & -58 to -32 & RCW~106  & & 881.48 & 85 & 2 & 2.0$\times10^{5}$ & 227.99 &50.02 & 20 & 500 \\
8 & 333.7 & -0.4 & C\hii & -80 to -57 & RCW~106 & 2.1 & 36.04 & 0.23 & 0.61 &
2.7$\times10^{3}$ & 74.15 & 47.29 & 0 & 0 \\

9 & 333.3 & -0.4 & \hii & -93 to -77 &  \gthree & 16.6 & 496.31 & 38 & 2.2 &
2.2$\times10^{4}$ & 44.45 & 49.65 & - & - \\

10 & 333 & -0.4 & \hii & -76 to -36 & RCW~106 & 19.6 & 402.26 & 73 & 2.6 &
1.2$\times10^{4}$ & 28.77 & 49.94 & 17 & 425 \\

11 & 333 & -0.7 & \hii & -76 to -36 & RCW~106 & 12.9 & - & 4.9 & 0.57 & - & - & 48.74 & 1 & 25 \\

12 & 332.9 & 0.8 & C\hii & -80 to -57 & RCW~106 & 5.6 & - & 6.6 & 2.1 & - & - & 48.93 & 1 & 25 \\

13 & 332.7 & -0.6 & \hii & -60 to -30 & RCW~106 & 16.8 & 103.62 & 36 & 1.6 &
1.0$\times10^{4}$ & 99.17 & 49.65 & 8 & 200 \\

14 & 332.5 & -0.1 & C\hii & -106 to -76 & \gthree & 4.3 & 83.84 & 1.9 & 1.1 &
9.8$\times10^{3}$ & 117.17 & - & - & - \\

& & & & -72 to -32 & RCW~106 & & 49.02 & & & 3.6$\times10^{3}$ & 73.84 &48.37 &0 & 0 \\

15 & 332.4 & 0.5 & SNR & - & - & 14.6 & 644.09 & 18 & 0.96 &
5.8$\times10^{4}$ & 90.73 & 49.4 &- & - \\

16 & 332.4 & -0.4 & SNR & - &- & 9.7 & - & 20 & 1.9 & - & - & 49.41 &
- & - \\

17 & 332.2 & -0.1 & C\hii & -101 to -77 & \gthree & 1.8 & 49.64 & 0.12 & 0.53 &
6.4$\times10^{3}$ & 129.6 & - & - & - \\

& & & & -77 to -48 &RCW~106  & & 32.21 & & & 1.6$\times10^{3}$ & 50.44 & 47.2 & 0 & 0 \\

18f1 & 333.1 & 0 & C\hii & -98 to -82 & \gthree  & 3.9 & 748.75 & 1.6 & 1.4 &
6.0$\times10^{4}$ & 80.66 & 48.26 & - & - \\

& & & & -73 to -36 & RCW~106  & & 469.51 & & & 3.8$\times10^{4}$ & 81.75 & 48.26 & 0 & 0 \\

18f2 & - & - & C\hii & -60 to -30  & RCW~106& 7.9 & - & 1.9 & 0.49 & - & - & 48.98 & 1 & 25 \\

18f3 & - & - & \hii & -60 to -30 & RCW~106 & 10.9 & - & 7.3 & 1.01 & - & - & 48.39 & 0 & 0 \\

19 & 332.1 & -0.4 & C\hii & -61 to -40 & RCW~106 & 7.2 & 65.68 & & & 2.5$\times10^{3}$ & 38.61 & -& - & - \\

20 & 331.9 & 0.2 & SNR & - &- & 11.1 & 368.16 & 3.7 & 0.56 &
3.3$\times10^{4}$ & 89.6 & 48.78 & - & - \\

21 & 331.6 & 0 & SNR & - &-& 13.6 & 1097.25 & 41 & 2.4 &
1.8$\times10^{5}$ & 166.59 & 49.71 & - & - \\

22 & 331.4 & -0.3 & \hii & -107 to -88 & \gthree & 9.7 & 407.17 & 7.9 & 1.5 &
4.3$\times10^{4}$ & 105.16 & - &- & - \\

& & & & -74 to -36 & RCW~106 & 216.67 & & & & 3.1$\times10^{4}$ & 144.96 & 48.97 &1 & 25 \\

23 & 331.3 & 0.5 & C\hii & -61 to -40 & RCW~106 & 8.2 & - & 2.4 & 0.59 & - & - & 48.59 & 0 & 0 \\

24 & 331.3 & 0 & C\hii & -104 to -72 &\gthree & 7.5 & 154.31 & 6.3 & 1.3 &
2.6$\times10^{4}$ & 169.12 &  - & - & - \\

& & & & -72 to -37 &RCW~106& & 96.66 & & & 8.2$\times10^{3}$ & 84.42 & 48.91 & 1 & 25 \\

25 & 331.2 & -0.2 & C\hii & -104 to -79 & \gthree & 4.9 & 123.88 & 3.4 & 1.7 &
2.3$\times10^{4}$ & 183.98 & - & - & - \\

& & & & -68 to -32 & RCW~106 & 69.47 & & & & 2.2$\times10^{3}$ & 32.31 & 48.63 & 0 & 0 \\

26 & 331.1 & -0.5 & C\hii & -68 to -32 & RCW~106 & 3.6 & - & 0.98 & 0.69 & - & - & 48.01 & 0 & 0 \\
26B & - & - & C\hii & -68 to -32 & RCW~106 & 3.7 & - & 1.1 & 0.84 & - & - & 48.14 & 0 & 0 \\

26C & - & - & C\hii & -68 to -32 & RCW~106  & 5.5 & - & 3.4 & 1.3 & - & - & 48.64 & 0 & 0 \\

27 & 331.1 & -0.2 & \hii & -101 to -76 & \gthree & 12.3 & 481.05 & 9.3 & 0.95 &
8.3$\times10^{4}$ & 171.78 & - & - & - \\

& & & & -69 to -33 & RCW~106 & 286.99 & & & & 1.6$\times10^{4}$ & 54.77 & 49.09 & 2 & 50 \\

28A & 330.7 & -0.4 & \hii & -101 to -80 & \gthree & 11.4 & 528.18 & 11 & 1.3 &
2.9$\times10^{4}$ & 55.97 & 49.15 &- & - \\

& & & & -69 to -32 & RCW~106 & 332.42 & & & & 4.1$\times10^{4}$ & 122.17 & 49.15 & 2 & 50 \\

28B & 330.7 & -0.4 & C\hii &  -69 to -32  & RCW~106& 3.9 & - & 1.8 & 1.4 & - & - & 47.54 & 0 & 0 \\
28C & 330.7 & -0.4  & C\hii &  -69 to -32  & RCW~106& 2.9 & - & 0.27 & 0.56 & - & - & 48.37 & 0 & 0 \\

29 & 330.3 & -0.4 & C\hii & -80 to -30 & RCW~106 & 2.8 & 13.68 & & & 1.2$\times10^{3}$ & 88.77 & 48.37 & 0 & 0 \\
30 & 330.2 & 1 & SNR & - & - & 5.9 & 244 & 1.3 & - & - & - & - & - & - \\

31 & 330.1 & -0.1 & C\hii & -112 to -80 & \gthree & 3.2 & 45.74 & - & - &
5.3$\times10^{3}$ & 115.8 & - & - & - \\

& & & & -67 to -33 & RCW~106 & & 23.44 & 0.67 & 0.86 & 1.0$\times10^{3}$ & 43.74 & 47.89 & 0 & 0 \\
\noalign{\vskip 2pt}
\hline
\noalign{\vskip 5pt}\end{tabular}

\end{threeparttable}
\label{box_table} 
\end{table*}

Across the continuum image the background is rather similar, about
$\sim 0.09$ Jy beam$^{-1}$, and contours are given starting at the $3
\sigma$ noise level above the local background.  We integrated the
background-subtracted 21~cm intensity within the area defined by the
3$\sigma$ noise contour.  The flux densities range from 0.1 to
12~Jy and the sizes range from a few arcminutes to several hundred
arcminutes.  There is no apparent correlation of size and flux
density. From these parameters and the distance, we have derived and
tabulated the characteristics of the star formation in
Table~\ref{box_table}.  Star formation in the RCW~106 complex is
discussed in Section~\ref{sect:sf}.

Because of the massive star formation across the region there are many
\hii regions.  These can be classified according to their shapes:
circular or irregular.  The \hii region around RCW~106 is classified
separately as a giant \hii region and the components are displayed in Figure~\ref{fig:rcw}.
There are also several SNRs seen in projection on
the region.
These structures can be seen in magnified views of these ``boxes" in
Figs.~\ref{fig:circ}--\ref{fig:supers} in the subsections below.
Within each figure, the boxes are shown on the same angular scale, as
labeled, with the box number at the lower left of each sub-image.  The
spatial locations and relationships of the boxes can be seen in
Fig.~\ref{21cm_box}.  In all figures, except Fig.~\ref{fig:supers} for
SNRs, a linear scale is also given, assuming a distance of 3.6~kpc,
although the structures associated with \gthree
(Table~\ref{box_table})
are at 5~kpc.

\newpage
\subsection{RCW~106: a giant \hii Region structure} \label{RCW_app}

\begin{figure}
\centering
\includegraphics[scale=0.41]{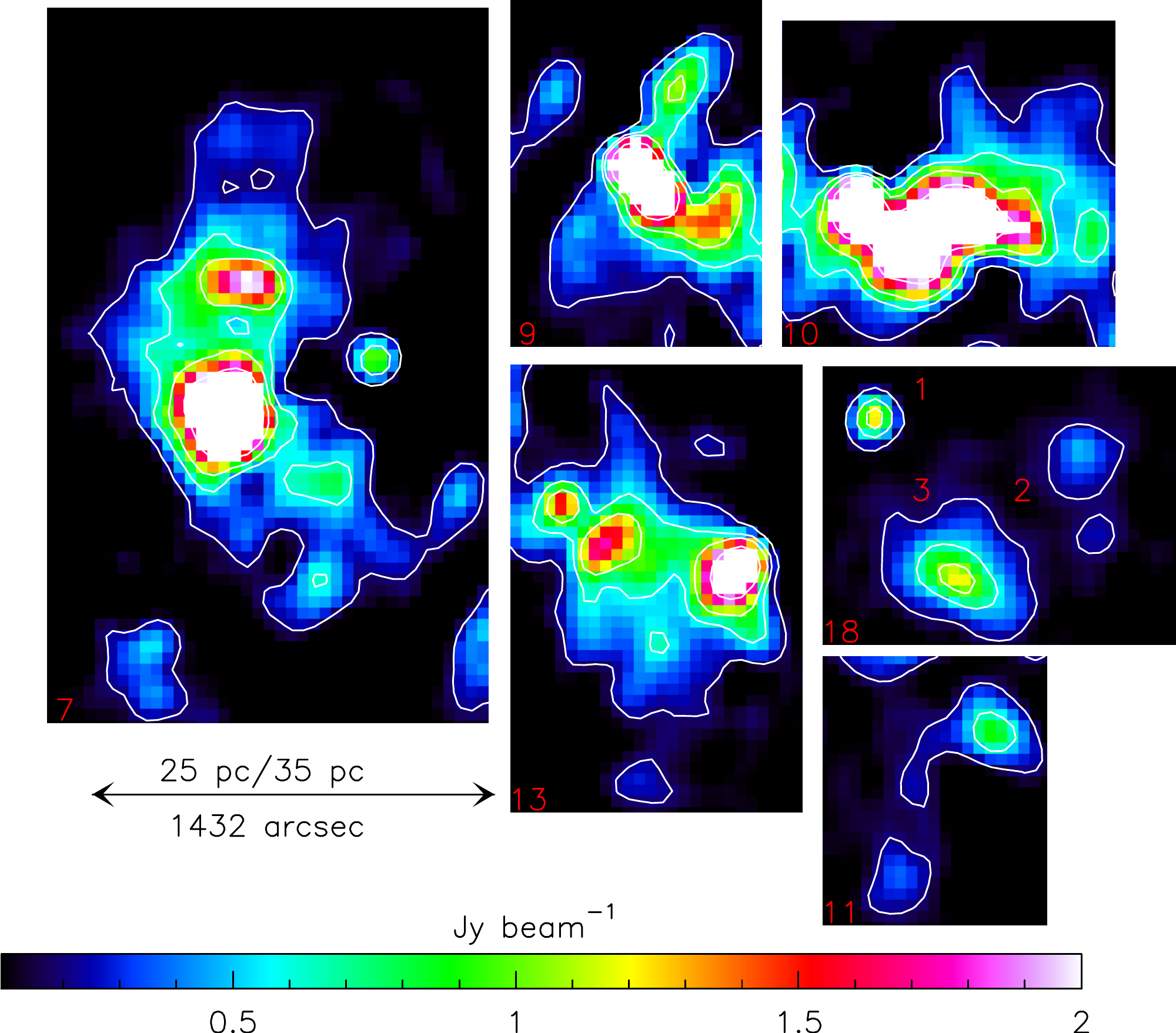}
\caption{Components of the giant \hii structure RCW~106.   The distance scales use both near and far kinematic distances.
\smallskip }
\label{fig:rcw}
\end{figure}

The \hii region complex RCW~106
spans approximately $1\fdeg5$ in the range $332\degr \le l \le
333\fdeg5$ {\bfc and so at a distance of 3.6 kpc  \citep{lockman79} is of ``giant" extent, 94~pc.}
The structure lies predominately below the Galactic
mid-plane.
{\bfc We separated the structure into six components which are large themselves}, illustrated in
Fig.~\ref{fig:rcw} and described in the brief notes below.

Box 7 (RCW106a). Large in size at nearly 33~pc. It has many known \hii
regions, star clusters, and SNR G333.6--00.2 which is located at the
brightest peak (see Fig. \ref{fig:rcw}a).

Box 9 (RCW106b). An extended structure of $\sim17$~pc in length (see
Fig.~\ref{fig:rcw}b). It has an irregular shape with prominent
protrusions. The main intensity peak is crescent shaped.

Box 10 (RCW106c). An irregular \hii region with a fairly large main
peak and size of almost 9~pc (see Fig.~\ref{fig:rcw}c).

Box 11 (RCW106d).  An irregular \hii region with a fairly large main
peak and a tail (see Fig.~\ref{fig:rcw}d).

Box~13 (RCW106e). An irregular structure with three main peaks loosely
connected and a size of about 17.5~pc (Fig.~\ref{fig:rcw}e).

Box 18 (RCW106f). Diffuse \hii region with four dominant peaks
(Fig.~\ref{fig:rcw}f).

\subsection{Circularly shaped \hii regions}\label{app:circ}

\begin{figure}
\centering
\includegraphics[width=8cm]{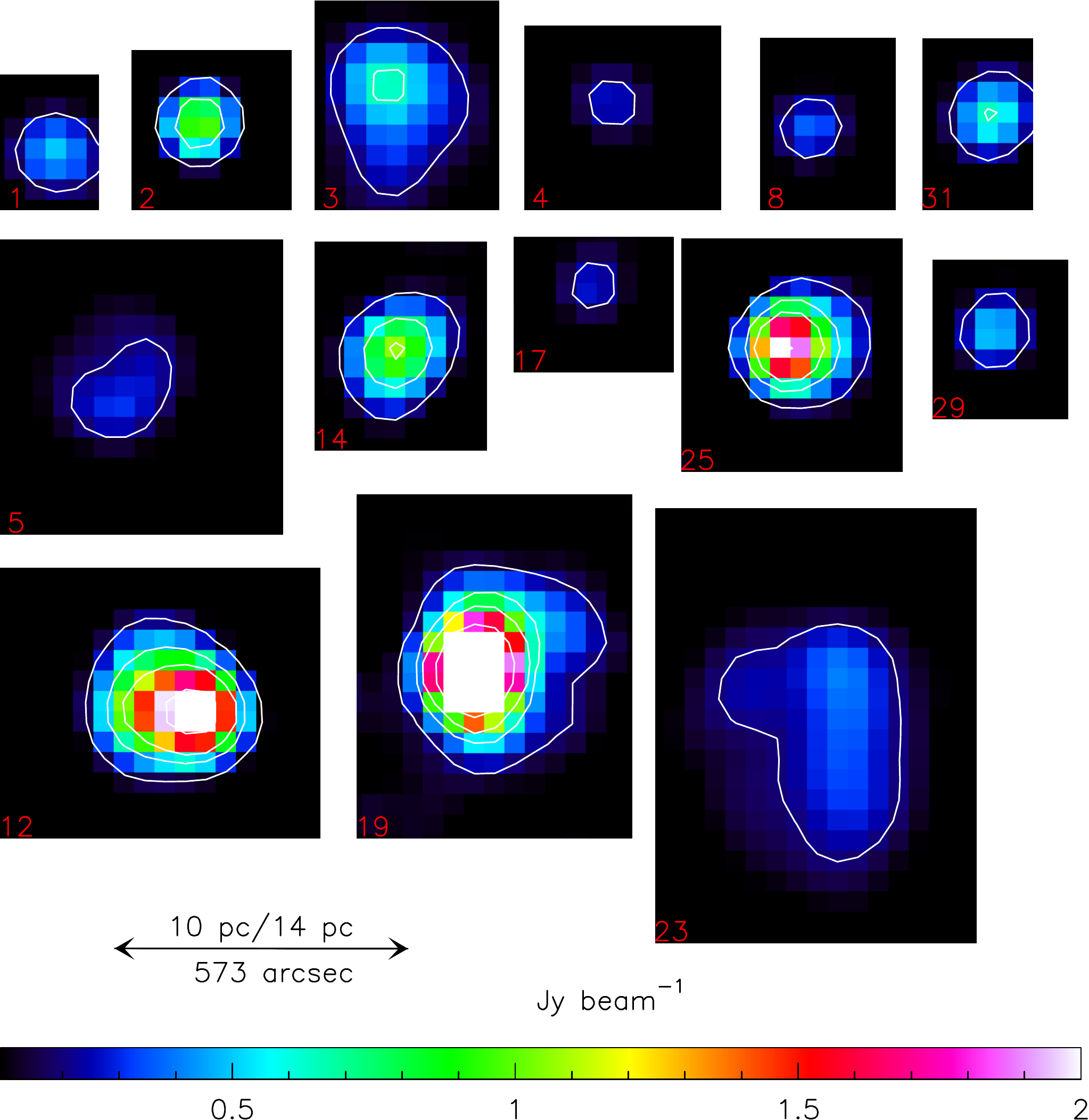}
\caption{Circularly shaped \hii regions. The distance scales use both near and far kinematic distances.}
\label{fig:circ}
\end{figure}

A number of the smaller \hii regions in Fig.~1 {\bfc appear to be fairly ``circular" in structure.}
These are illustrated in Fig.~\ref{fig:circ} and described
below.

Box 1. The compact \hii region contains the single IRAS source
16226--4900 at its center.

Box 2.  An ultracompact \hii region coincides with an IR star cluster
[DBS 2003] 170.\footnote{ [\emph{ref.}]\,\# from SIMBAD:
  \url{http://simbad.u-strasbg.fr}.}  As discussed by \cite{dutra03},
the composition is a bit ambiguous but there are a few bright
stars. The \hii region GAL334.71--00.67 found in this box is 17.4 kpc
away as determined by its positive velocity \citep{russeil05}.

Box 3. The compact \hii region that we detect is roughly 5.8~pc in
size and probably coincides with the \hii region GAL 334.68--00.11. It
has a nearby cluster of IR sources \citep{robitaille08}.

Box 4. This compact \hii region is roughly 1.6~pc in size, which almost
makes it an ultracompact candidate.

Box 5. There is a YSO on the border of the compact \hii region. The
star cluster [MCM 2005b] 79 \citep{mcm05b} is also nearby.
{\bfc This source has a faint protrusion in the lowest contour, not so significant that we describe it as
  irregular.}

Box 8. This compact \hii region has a few YSOs nearby.

Box 12. This compact \hii region contains the IR star cluster
[DBS 2003] 102. It also coincides with the known \hii regions GAL
332.98+00.79 and [WHR97] 16112--4943. The latter is classified as
UCHII \citep{walsh97}.

Box 14. This compact grouping encompasses the known \hii regions
[KC 97c] G332.5--00.1, GRS 332.54--00.11, and GAL 332.54--00.11
\citep{kc97c} and the star clusters [DBS 2003] 160 and 161. This clump
is a potential OB cluster as seen from the numerous OB stars within
the region. There are also some YSOs and outflow regions within the
clump.

Box 17. This is the compact \hii region \emph{IRA}S 16119-5048 driving an
outflow.

Box 19. An \hii region hosting a star cluster [MCM 2005b] 74. It has a
cometary shape and is significantly brighter with an off-center peak
in its circular structure. It is located near the supernova remnant,
SNR G332.4-0.4.

Box 23. Most likely the compact \hii region GAL 331.36+00.51, the
structure is elliptical with a slight protrusion along the
longitudinal axis and is 8.6 pc in size.

Box 25.  A possible compact \hii region. It is $\sim 5$~pc in size
with the known \hii regions [KC 97c] G331.3--00.2, GAL 331.28--00.19,
and GAL 331.26--00.19 within its peak.

Box 29. A compact \hii region, 2.9~pc in size and associated with the
IR cluster [DBS 2003] 153. It also hosts the know \hii region \emph{IRAS}
16037--5223 and GAL 330.31--00.39.

Box 31. A compact \hii region probably powered by GAL~330.04--00.05.

\subsection{Irregularly shaped \hii regions}
\label{app:irreg}

\begin{figure}[t]
\centering
\includegraphics[width=8cm]{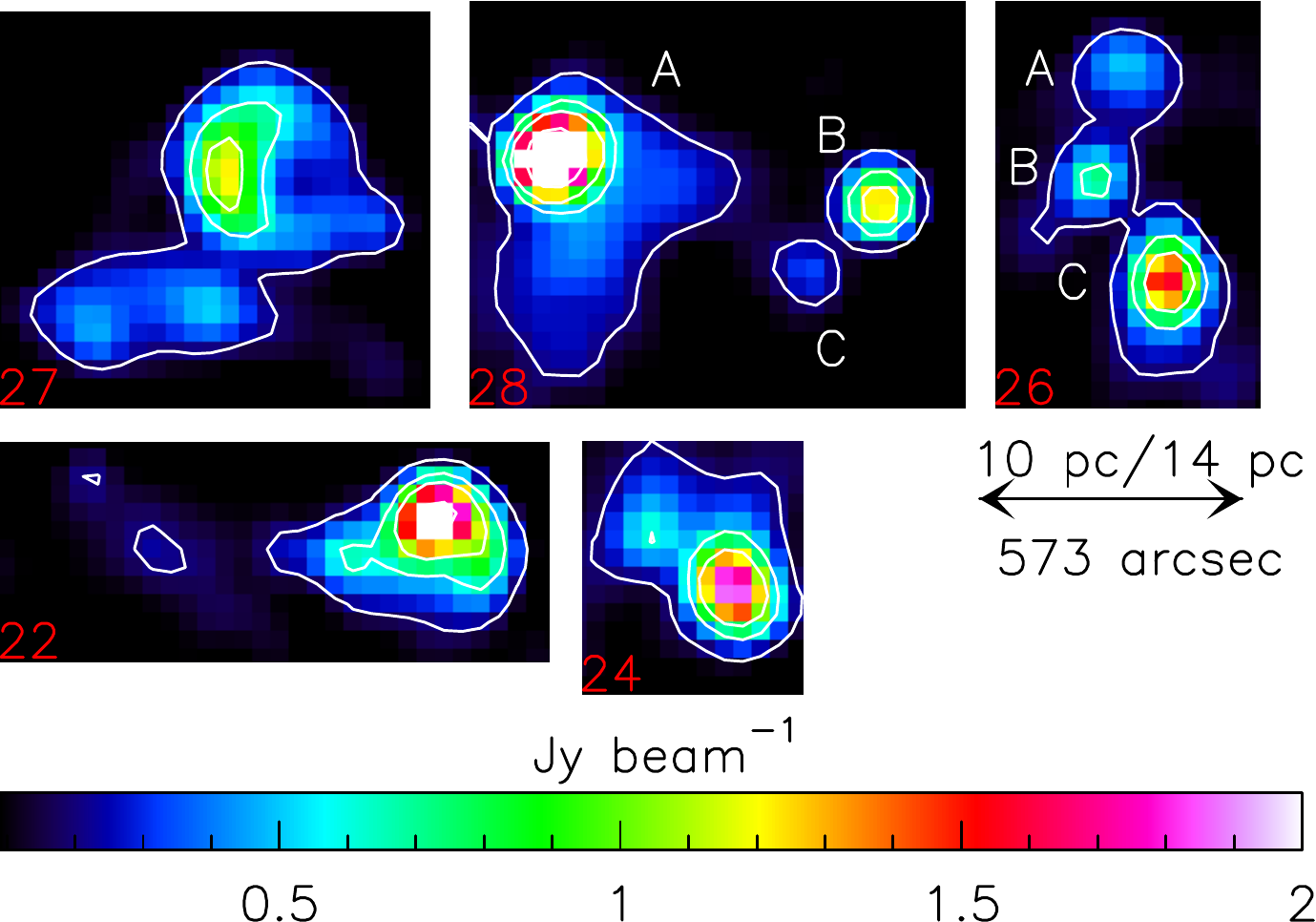}
\caption{Irregularly shaped \hii regions. The distance scales use both near and far kinematic distances.}
\label{fig:irreg}
\end{figure}

The remaining \hii regions as seen in
Fig.~\ref{fig:irreg} are {\bfc ``irregular" in outline as compared those in Fig.~\ref{fig:circ}.}
Descriptions of these follow.

Box 22. An extended \hii region of $\sim 10$ pc in size. It has
numerous IR star clusters and known \hii regions. It displays a
comet-like structure with its brightest and largest feature at
$331\fdeg33$. There are two smaller peaks trailing off from the
tail. They are diffuse and seemingly attached to the main structure.

Box 24. A compact \hii region with the IR star cluster, [DBS 2003]
158. It is elongated and has double peaks. The main peak is off-center
in one of the protrusions where the known \hii regions are located. It
is also near the supernova candidate MSC331.8+0.0 whose structure may
be related.

Box 26. The structure is elongated with three main compact peaks that
may all be connected. Together they form an elongated structure that
is $\sim 11$ pc in length.

Box 27. Both GAL 331.03--00.15 and PMN J1610--5150 coincides with this
\hii region. There are numerous OB associations spread around the main
intensity peak.

Box 28. This has been subdivided into three regions 28A, 28B, and 28C
from largest to smallest with sizes $\sim 12$, 4, and 3~pc,
respectively. The \hii region 28A hosts numerous known \hii regions
and the IR star cluster [DBS 2003] 155 while 28B appears as a compact
\hii region with IR star cluster [DBS 2003] 154. 28C is as yet
undefined. Individually, 28A has a comet-like shape with two tails
while 28B and 28C are circular in nature.

\subsection{Supernova remnants and candidates}\label{app:snr}

\begin{figure}[b]
\centering
\includegraphics[width=8cm]{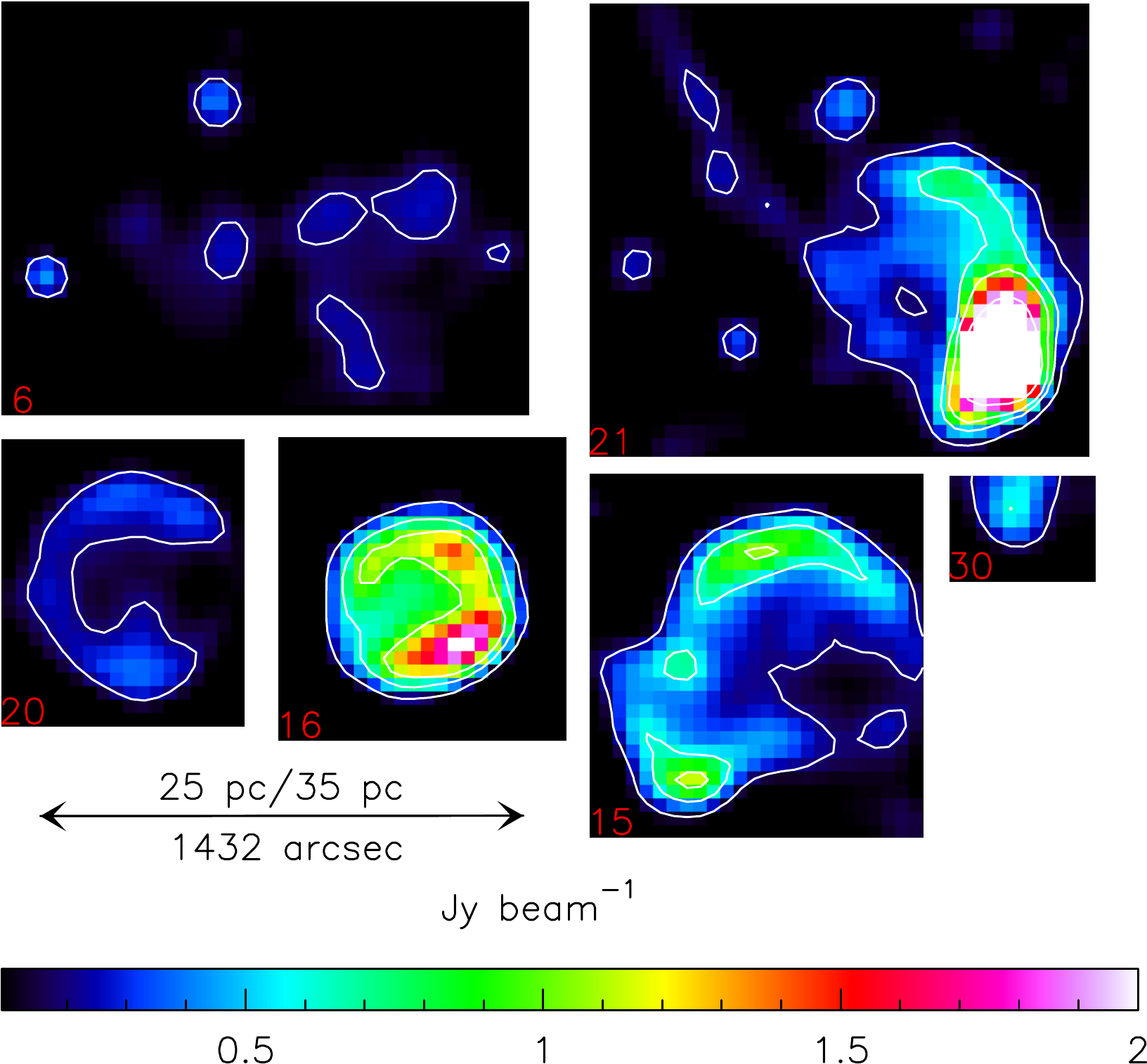}
\caption{Supernova remnants and candidates. The distance scales use both near and far kinematic distances.}
\label{fig:supers}
\end{figure}

There are four confirmed and two candidate supernova remnants (SNRs)
(Fig.~\ref{fig:supers}).  These are seen in projection and are not
necessarily associated with the two main star-forming complexes.

Box 6. The SNR Candidate G333.9+00.0 seems very faint with many separated clumps.

Box 15. The SNR G332.4+0.1 has three main peaks that may be part of one
shell judging from the contour lines. From OH absorption it is thought
to be in the range of 7--11~kpc \citep{caswell75}.

Box 16. The SNR G332.4--0.4 seems to have a circular shell. A higher
intensity portion situated near the bottom is the RCW~103 star
cluster. It has a kinematic distance of 3.1~kpc from \hi absorption
lines \citep{green09}.

Box 20. The shell of SNR G332.0+0.2 is a crescent and is thought to be
at at least 7~kpc \citep{caswell75}.

Box 21. The SNR candidate MSC 331.8+0.0 shows a high intensity portion
with scattered filaments.
Box 30. The SNR G330.2+01.0 has a minimum distance of 4~kpc \citep{mcclure01}.

\end{document}